\documentclass[prd,twocolumn,superscriptaddress,floatfix,nofootinbib]{revtex4-2}

\usepackage[T1]{fontenc}
\usepackage{amsmath}
\usepackage{amssymb}
\usepackage{amsfonts}
\usepackage{graphicx}
\usepackage{enumitem}
\usepackage{bm}
\usepackage{rotating}
\usepackage{hyperref}
\usepackage{pbox}
\usepackage[dvipsnames]{xcolor}
\usepackage{array}
\usepackage{physics}
\usepackage{dsfont}
\usepackage{mathtools}
\usepackage{leftidx}
\usepackage{lmodern}
\usepackage[normalem]{ulem}
\usepackage{braket}
\usepackage{tensor}
\usepackage{mathrsfs}
\usepackage{float}
\usepackage{dsfont}
\usepackage[margin=2cm]{geometry}
\usepackage[title]{appendix}
\usepackage{tcolorbox}

\usepackage{tikz}
\usetikzlibrary{arrows.meta,decorations.pathmorphing,math}

\renewcommand{\Re}{\mathrm{Re}}
\renewcommand{\Im}{\mathrm{Im}}
\renewcommand{\bar}{\overline}
\renewcommand{\tilde}{\widetilde}
\newcommand{\skri}{\mathscr{I}}
\newcommand{\ii}{\mathsf{i}}
\newcommand{\hor}{\mathscr{H}}
\newcommand{\sx}{\mathsf{x}}
\newcommand{\sy}{\mathsf{y}}

\newcommand{\rr}[1]{\left(#1\right)}
\newcommand{\bx}{{\bm{x}}}
\newcommand{\by}{{\bm{y}}}
\newcommand{\bk}{{\bm{k}}}
\newcommand{\br}{{\bm{r}}}
\newcommand{\sK}{\mathsf{K}}
\newcommand{\supp}{\text{supp}}

\newcommand{\VS}{{\mathfrak{X}}}

\newcommand{\R}{\mathbb{R}}
\newcommand{\C}{\mathbb{C}}
\newcommand{\M}{\mathcal{M}}
\newcommand{\N}{\mathcal{N}}
\newcommand{\A}{\mathcal{A}}
\newcommand{\W}{\mathcal{W}}
\newcommand{\CS}{C^\infty_0(\M)}
\newcommand{\Sol}{\mathsf{Sol}}

\begin{document}

\title{Modest holography and bulk reconstruction in asymptotically flat spacetimes}

\author{Erickson Tjoa}
\email{e2tjoa@uwaterloo.ca}
\affiliation{Department of Physics and Astronomy, University of Waterloo, Waterloo, Ontario, N2L 3G1, Canada}
\affiliation{Institute for Quantum Computing, University of Waterloo, Waterloo, Ontario, N2L 3G1, Canada}

\author{Finnian Gray}
\email{fgray@perimeterinstitute.ca}
\affiliation{Department of Physics and Astronomy, University of Waterloo, Waterloo, Ontario, N2L 3G1, Canada}
\affiliation{Perimeter Institute for Theoretical Physics, Waterloo, Ontario N2L 2Y5, Canada}

\begin{abstract}

In this work we present a ``modest'' holographic reconstruction of the bulk geometry in asymptotically flat spacetime using the two-point correlators of boundary quantum field theory (QFT) in asymptotically flat spacetime. The boundary QFT lives on the null boundary of the spacetime, namely null infinity and/or the Killing horizons. The bulk reconstruction relies on two unrelated results: (i) there is a bulk-to-boundary type correspondence between free quantum fields living in the bulk manifold and free quantum fields living on its null boundary, and (ii) one can construct the metric by making use of the Hadamard expansion of the field living in the bulk. This holographic reconstruction is ``modest'' in that {the fields used are non-interacting and not strong-weak holographic duality in the sense of AdS/CFT, but it  works for generic asymptotically flat spacetime subject to some reasonably mild conditions.}

\end{abstract}

\maketitle
\flushbottom

\section{Introduction}

In general relativity, more often than not any reasonable observers are located far away from any astrophysical objects, thus in many situations one can approximate observers as essentially at infinity. This is especially evident in the detection of electromagnetic and gravitational radiation from some astrophysical sources. At the same time, electromagnetic and gravitational radiations travel along null geodesics, thus they will reach future \textit{null infinity} $\skri^+$. No observers can be exactly at $\skri^+$, but for many practical calculations one can approximate them to be ``close'' to null infinity to detect these radiations. Therefore, physics at null infinity remains very relevant for studying what faraway observers can see.

One less well-known but nonetheless remarkable result in algebraic quantum field theory (AQFT) is that there is a form of \textit{bulk-to-boundary correspondence} between massless QFT living in the bulk geometry and massless QFT living in its null boundary \cite{Dappiaggi2005rigorous-holo,Dappiaggi2008cosmological,Dappiaggi2009Unruhstate,dappiaggi2015hadamard}. In the case of asymptotically simple spacetimes (i.e., without horizons), the null boundary is simply the null infinity, while for Schwarzschild geometry this will be the union of the Killing horizon and null infinity. This provides a form of holography between the algebra of observables  and states of two scalar field theories. However, this is arguably less attractive compared to the holographic duality provided by Anti-de Sitter/Conformal Field Theory (AdS/CFT) correspondence \cite{Maldacena1999holography,witten1998anti,Gubser1998gaugestring} (see, e.g., \cite{Hubeny2015review,Aharony2000largeN,horowitz2009gauge,RyuTakayanagi2006,akers2020simple} and references therein for non-exhaustive list of this very vast research program). There, the CFT can be very strongly coupled and it can be used to construct \textit{directly} the bulk asymptotically AdS geometry. The flat holography presented above is really about reconstruction of \textit{correlators} of the bulk \textit{non-interacting} QFT from the correlators of another non-interacting QFT at the boundary. As such, they give us a very different and modest kind of holography, as was already pointed out in \cite{Dappiaggi2005rigorous-holo}.

In this work, we will show that we can do better by actually reconstructing the metric of the bulk geometry directly from the boundary correlators (the smeared $n$-point functions). We use a method inspired from the work of Saravani, Aslanbeigi, and Kempf  \cite{Kempf2016curvature,Kempf2021replace}, where they reconstructed the bulk metric from bulk scalar propagators (the Feynman propagator). The basic idea is that physically reasonable states fall under the class of \textit{Hadamard states} \cite{Khavkhine2015AQFT,KayWald1991theorems,Radzikowski1996microlocal}, which have the property that the short-distance (UV) behaviour of the correlation functions is dominated by the geodesic distance between two nearby points (typically written in terms of Synge world functions). By augmenting the bulk-to-boundary correspondence in \cite{Dappiaggi2005rigorous-holo,Dappiaggi2008cosmological,Dappiaggi2009Unruhstate,dappiaggi2015hadamard} with the metric reconstruction scheme in \cite{Kempf2016curvature,Kempf2021replace} but replacing the bulk propagator with the boundary correlation functions, we will be able to establish a form of holographic bulk reconstruction. This works because the Hadamard property of the states in the bulk is encoded non-trivially into the boundary correlators. Note that the metric reconstruction from bulk Wightman two-point functions, exploiting the Hadamard property directly, was done explicitly for the first time in \cite{perche2021geometry}. We will show this reconstruction using Minkowski and Friedmann-Robertson-Walker (FRW) spacetimes.  We will refer to this version of bulk-to-boundary correspondence ``modest holography''.
 
We should emphasize what we are \textit{not} doing in this work. We do not claim that we can reconstruct \textit{all} asymptotically flat spacetimes purely from $\skri^+$, and certainly not  the maximal analytic extensions in general. The modest holography works as far as there is enough ``Cauchy data'' {at $\skri^+$} for the reconstruction. For example, if we have a black hole with future horizon $\mathscr{H}^+$, observers near $\skri^+\cup\mathscr{H}^+$ can {at most} reconstruct the metric holographically in the \textit{exterior} of the black hole. The reason is simply that there is not enough Cauchy data to reconstruct the interior using this method. {In some cases one may need to include timelike infinity (even for massless fields) to have enough Cauchy data~\cite{Geroch:1978us}. Note also that violation of strong Huygens' principle in generic curved spacetimes means that massless field causal propagators can have timelike supports \cite{Valerio1999tails,Sonego1992huygens}. {In this respect, our bulk reconstruction construction is closer to that of Hamilton-Kabat-Lifschytz-Lowe (HKLL) construction in AdS/CFT \cite{Bena2000local,Hamilton2006,Hamilton2006local2}}. } What we propose here is that the bulk-to-boundary correspondence proposed in \cite{Dappiaggi2005rigorous-holo,Dappiaggi2008cosmological,Dappiaggi2009Unruhstate,dappiaggi2015hadamard}, which was only between bulk and boundary correlators, can (and perhaps should) be promoted to an actual holographic reconstruction of the bulk geometry. The limitation of the bulk reconstruction is ultimately {dependent on} to the validity of the modest holography itself.

It is worth mentioning that our results are only guaranteed in (3+1)-dimensional asymptotically flat spacetimes, where the asymptotic symmetry group is the \textit{Bondi-Metzner-Sachs (BMS) group} \cite{bondi1962gravitational,sachs1962gravitational}. In higher dimensions this may not be the case and it has been debated in the literature when the BMS group remains the asymptotic symmetry group (see, e.g., \cite{Hollands2017higherDimBMS,Pate2018higherDimBMS}). This is closely tied to the existence of gravitational memory effect. We are not aware of any analogous bulk-to-boundary correspondence in higher dimensions.

A side goal of this work is to popularize the technique in AQFT in more accessible manner to people working in QFT in curved spacetimes and also the RQI community. One of the authors provided similar introduction to AQFT in \cite{tjoa2022channel}, and in this work we will refine some of the exposition, and complement it by also providing an accessible introduction to the algebraic framework for scalar QFT on $\skri^+$. Notation-wise we will combine the best features of \cite{wald1994quantum,KayWald1991theorems,Landulfo2021cost,Moretti2005BMS-invar,Moretti2008outstates,Tales2020GRQO,tjoa2022channel,Khavkhine2015AQFT,fewster2019algebraic}. This work is an extension to the shorter work in \cite{tjoa2022holographic-essay}. {A much more extensive description of algebraic framework for fields of various spins and masses in the context of $S$-matrix formalism is given very recently in \cite{prabhu2022infrared}, which shares similar language with what we do here.}

This work is organized as follows. In Sec.~\ref{sec: AQFT} we will briefly review the algebraic framework for real scalar QFT in arbitrary globally hyperbolic spacetimes. In Section~\ref{sec: AQFT-null} we briefly review algebraic framework for real scalar QFT living on null infinity. In Section~\ref{sec: holographic-reconstruction} we present an explicit calculation for the holographic reconstruction of the bulk correlators from its boundary correlators and show how to construct the bulk metric. In Section~\ref{sec: large-r-expansion} we discuss the connection with large-$r$ expansion of the bulk fields. In Section~\ref{sec: discussion} we discuss our results and outlook for further investigations.

{\textit{Conventions:} we use the convention $c=\hbar=1$ and mostly-plus signature for the metric. Also, in order to match both physics and mathematics literature without altering each other's conventions, we will make the following compromise. In most places we will follow ``{physicist's convention}'', writing Hermitian conjugation as $A^\dagger$ and complex conjugation as $B^*$. There will be three exceptions using ``{mathematician's convention}'': (1) $C^*$-algebra in Section~\ref{sec: AQFT}, where $*$ here really means (Hermitian) adjoint/Hermitian conjugation (2) {complex conjugate Hilbert space} $\overline{\mathcal{H}}$ in Section~\ref{sec: AQFT}, and (3) complex stereographic coordinates $(z,\overline{z})$ in Appendix~\ref{appendix: BMS}, where complex conjugation is denoted by a bar.}

\section{Scalar QFT in curved spacetimes}
\label{sec: AQFT}

In this section we briefly review the algebraic framework for quantization of real scalar field in arbitrary (globally hyperbolic) curved spacetimes. {We will follow {largely} the conventions of Kay and Wald~\cite{KayWald1991theorems} with small modification\footnote{{This will be slightly different from the conventions used by one of us in \cite{tjoa2022channel} which is closer to \cite{Landulfo2016magnus1,Landulfo2021cost}.}} We caution the reader that in the AQFT literature there are various different conventions being used (\textit{c.f.} \cite{wald1994quantum,Dappiaggi2005rigorous-holo,Moretti2005BMS-invar,Khavkhine2015AQFT,Landulfo2021cost,fewster2019algebraic}), in particular the convention regarding about symplectic smearing (we explain some of these in Appendix~\ref{appendix: symplectic-smearing-Wald}).} {An accessible introduction to $*$-algebras and $C^*$-algebras can be found in \cite{fewster2019algebraic}.}

\subsection{Algebra of observables and algebraic states}

Consider a free, real scalar field $\phi$ in (3+1)-dimensional globally hyperbolic spacetime $(\mathcal{M},g_{ab})$. Global hyperbolicity guarantees that $\M$ admits foliation by spacelike Cauchy surfaces $\Sigma_t$ labelled by real (time) parameter $t$. The field generically obeys the Klein-Gordon equation
\begin{align}
     P\phi = 0\,,\quad  P = \nabla_a\nabla^a - m^2  - \xi R\,,
     \label{eq: KGE}
\end{align}
where $\xi \geq 0$, $R$ is the Ricci scalar and  $\nabla$ is the Levi-Civita connection with respect to $g_{ab}$. Later we specialize to massless conformally coupled fields.

Let $f\in \CS$ be a smooth compactly supported test function on $\M$. Let $E^\pm(\sx,\sy)$ be the retarded and advanced propagators associated to the Klein-Gordon operator $P$, such that
\begin{align}
    E^\pm f\equiv (E^\pm f)(\sx) \coloneqq \int \dd V'\, E^\pm (\sx,\sx')f(\sx') \,,
\end{align}
solves the inhomogeneous equation $P(E^\pm f) = f$. Here $\dd V' = \dd^4\sx'\sqrt{-g}$ is the invariant volume element. The \textit{causal propagator} is defined to be the advanced-minus-retarded propagator $E=E^--E^+$. If $O$ is an open neighbourhood of some Cauchy surface $\Sigma$ and $\varphi$ is any real solution with compact Cauchy data to Eq.~\eqref{eq: KGE}, which we denote by $\varphi \in \Sol_\R(\M)$, then there exists $f\in \CS$ with $\supp(f)\subset O$ such that $\varphi=Ef$ \cite{Khavkhine2015AQFT}.

Let us now review the algebraic approach to free, real scalar quantum field theory {(see the comparison with canonical quantization formulation in Appendix A of \cite{tjoa2022channel}, also \cite{fewster2019algebraic,Khavkhine2015AQFT,KayWald1991theorems}).} In AQFT, the quantization of real scalar field is regarded as an $\R$-linear mapping from the space of smooth compactly supported test functions to a unital $*$-algebra $\A(\M)$
\begin{align}
    \hat\phi: C^\infty_0(\mathcal{M})&\to \A(\M)\,,\quad f\mapsto \hat\phi(f)\,,
\end{align}
which obeys the following conditions:
\begin{enumerate}[leftmargin=*,label=(\alph*)]
    \item (\textit{Hermiticity}) $\hat\phi(f)^\dag = \hat\phi(f)$ for all $f\in \CS$;
    \item (\textit{Klein-Gordon}) $\hat\phi(Pf) = 0$ for all $f\in \CS$;
    \item (\textit{Canonical commutation relations}  (CCR)) we have $[\hat\phi(f),\hat\phi(g)] = \ii E(f,g)\openone $ for all $f,g\in \CS$, where $E(f,g)$ is the smeared causal propagator\footnote{We have removed the redundant notation $\Delta(f,g)$ in \cite{Landulfo2016magnus1,Landulfo2021cost,tjoa2022channel}.}
    \begin{align}
        E(f,g)\coloneqq \int \dd V f(\sx) (Eg)(\sx)\,.
    \end{align}
    \item (\textit{Time slice axiom}) Let $\Sigma\subset \M$ be a Cauchy surface and $O$ a fixed open neighbourhood of $\Sigma$. $\A(\M)$ is generated by the unit element $\openone$ (hence $\A(\M)$ is unital) and the smeared field operators $\hat\phi(f)$ for all $f\in \CS$ with $\supp(f)\subset O$.
\end{enumerate}
The $*$-algebra $\A(\M)$ is called the \textit{algebra of observables} of the real Klein-Gordon field. The \textit{smeared} field operator reads
\begin{align}\label{eq: ordinary smearing}
    \hat\phi(f) = \int \dd V\hat\phi(\sx)f(\sx)
\end{align}
and $\hat\phi(\sx)$ is to be regarded as an operator-valued distribution. 

The algebra of observables $\A(\M)$ defined above is still somewhat abstract. This can be made more concrete by making explicit the symplectic structure of the theory. First, the vector space of real-valued solutions of Klein-Gordon equation with compact Cauchy data, denoted $\Sol_\R(\M)$, can be made into a \textit{symplectic} vector space by equipping it with a symplectic form $\sigma:\Sol_\R(\M)\times\Sol_\R(\M)\to \R$, defined as
\begin{align}
    \sigma(\phi_1,\phi_2) \coloneqq \int_{\Sigma_t}\!\! {\dd\Sigma^a}\,\Bigr[\phi_{{1}}\nabla_a\phi_{{2}} - \phi_{{2}}\nabla_a\phi_{{1}}\Bigr]\,,
    \label{eq: symplectic form}
\end{align}
where $\dd \Sigma^a = -t^a \dd\Sigma$, $-t^a$ is the inward-directed unit normal to the Cauchy surface $\Sigma_t$, and $\dd\Sigma = \sqrt{h}\,\dd^3\bx$ is the induced volume form on $\Sigma_t$ \cite{Poisson:2009pwt,wald2010general}. This definition is independent of the Cauchy surface. With this, we can regard $\hat\phi(f)$ as \textit{symplectically smeared field operator}  \cite{wald1994quantum} 
\begin{align}
    \label{eq: symplectic smearing}
    {\hat\phi(f) \equiv \sigma(Ef,\hat\phi)\,,}
\end{align}
and the CCR algebra can be written as 
\begin{align}
    {[\sigma(Ef,\hat\phi),\sigma(Eg,\hat\phi)] = \ii\sigma(Ef,Eg)\openone = \ii E(f,g)\openone \,,}
\end{align}
where $\sigma(Ef,Eg) = E(f,g)$ in the second equality follows from Eq.~\eqref{eq: ordinary smearing} and \eqref{eq: symplectic smearing}. The symplectic smearing has the advantage of keeping the dynamical content manifest at the level of the field operators (via the causal propagator). We will see later that sometimes it is much more obvious how to proceed with this interpretation than working abstractly, especially so for scalar QFT at $\skri^+$. For convenience, we will collect some results involving symplectic smearing in Appendix~\ref{appendix: symplectic-smearing-Wald}.

In many cases, it is more convenient to work with the ``exponentiated'' version of $\A(\M)$ called the \textit{Weyl algebra} (denoted by $\W(\M)$), {since its elements are (formally) bounded operators.} The Weyl algebra $\W(\M)$ is a unital $C^*$-algebra generated by the elements which formally take the form 
\begin{align}
    W(Ef) \equiv 
    {e^{\ii\hat\phi(f)}}\,,\quad f\in \CS\,.
    \label{eq: Weyl-generator}
\end{align}
These elements satisfy \textit{Weyl relations}:
\begin{equation}
    \begin{aligned}
    W(Ef)^\dagger &= W(-Ef)\,,\\
    W(E (Pf) ) &= \openone\,,\\
    W(Ef)W(Eg) &= e^{-\frac{\ii}{2}E(f,g)} W(E(f+g))
    \end{aligned}
    \label{eq: Weyl-relations}
\end{equation}
where $f,g\in \CS$. The symplectic smearing picture has the advantage that even for the Weyl algebra $\W(\M)$, microcausality can be given in the same way as CCR algebra of $\A(\M)$. That is, using $\sigma(Ef,Eg) = E(f,g)$, the Weyl relations Eq.~\eqref{eq: Weyl-relations}, and the fact that $\supp(Ef)\subset J^+(\supp(f))$ where $J^+(\supp(f))$ is the causal future of $\supp(f)$, we have \cite{Dappiaggi2005rigorous-holo}
\begin{align}
    [W(Ef),W(Eg)] = 0 \,,
    \label{eq: Weyl-commutator-bulk}
\end{align}
whenever $\supp(f)\,\cap\,\supp(g) = \emptyset$ (supports of $f$ and $g$ are causally disjoint, i.e., ``spacelike separated'')\footnote{Abstractly, one would have considered elements of the Weyl algebra to be $W(\phi)$ for some $\phi\in \Sol_\R(\M)$. In this form, microcausality is far from obvious because the third Weyl relation would have read $W(\phi_1)W(\phi_2) = e^{{-}\ii\sigma(\phi_1,\phi_2)/2}W(\phi_1 + \phi_2)$. }.

After specifying the algebra of observables, we need to provide a quantum state for the field. In AQFT the state is called an \textit{algebraic state}, defined by a $\C$-linear functional $\omega:\A(\M)\to \C$ such that 
\begin{align}
    \omega(\openone) = 1\,,\quad  \omega(A^\dagger A)\geq 0\quad \forall A\in \A(\M)\,.
    \label{eq: algebraic-state}
\end{align}
That is, a quantum state is normalized to unity and positive-semidefinite operators have non-negative expectation values. The state $\omega$ is pure if it cannot be written as $\omega= \alpha \omega_1 + (1-\alpha)\omega_2$ for any $\alpha\in (0,1)$ and any two algebraic states $\omega_1,\omega_2$; otherwise the state is said to be mixed. 

The {connection to the usual notion of Hilbert spaces comes from the} Gelfand-Naimark-Segal (GNS) reconstruction theorem \cite{wald1994quantum,Khavkhine2015AQFT,fewster2019algebraic}. This says that we can construct a \textit{GNS triple}\footnote{Strictly speaking we also need to provide a dense subset $\mathcal{D}_\omega\subset \mathcal{H}_\omega$ since the field operators are unbounded operators.} $(\mathcal{H}_\omega, \pi_\omega,{\ket{\Omega_\omega}})$, where $\pi_\omega: \mathcal{\A(\M)}\to {\text{End}(\mathcal{H}_\omega)}$ is a Hilbert space representation with respect to state $\omega$ such that any algebraic state $\omega$ can be realized as a \textit{vector state} {$\ket{\Omega_\omega}\in\mathcal{H}_\omega$}. The observables $A\in \A(\M)$ are then represented as operators $\hat A\coloneqq \pi_\omega(A)$ acting on the Hilbert space. With the GNS representation, the action of algebraic states take the familiar form
\begin{align}
    \omega(A) = \braket{\Omega_\omega|\hat A|\Omega_\omega}\,.
\end{align}
The main advantage of the AQFT approach is that it is independent of the representations of the CCR algebra chosen: there are as many representations as there are algebraic states $\omega$. Since QFT in curved spacetimes admits infinitely many unitarily inequivalent representations of the CCR algebra, the algebraic framework allows us to work with all of them at once.

In the case of Weyl algebra, the algebraic state and GNS representation gives concrete realization of ``exponentiation of $\hat\phi(f)$''. The exponentiation in Eq.~\eqref{eq: Weyl-generator} is only formal: we \textit{cannot} literally regard the smeared field operator $\hat\phi(f)$ as the derivative $\partial_t\bigr|_{t=0}W(t Ef)$ since the Weyl algebra itself does not have the right topology \cite{fewster2019algebraic}; instead one takes the derivative at the level of the GNS representation: that is, if $\Pi_\omega:\W(\M)\to \mathcal{B}(\mathcal{H}_\omega)$ is a GNS representation with respect to $\omega$, then we do have 

\begin{align}
    \pi_\omega(\hat\phi(f)) &= -\ii\frac{\dd}{\dd t}\Bigg|_{t=0}\!\!\!\!\!\!\!\Pi_\omega(e^{\ii t \hat\phi(f)}) \equiv - \ii\frac{\dd}{\dd t}\Bigg|_{t=0}\!\!\!\!\!\!\!e^{\ii t \pi_\omega(\hat\phi(f))} \,,
\end{align}
where now $\hat\phi(f)$ is smeared field operator acting on Hilbert space $\mathcal{H}_\omega$. We can then \textit{define} the formal $n$-point functions to be the expectation value in its GNS representation. For example, in the case of two-point functions we have
\begin{align}\label{eq: Wightman-formal-bulk}
    &\omega \bigr(\hat\phi(f)\hat\phi(g)\bigr) \coloneqq \braket{\Omega_\omega|\pi_\omega(\hat\phi(f))\pi_\omega(\hat\phi(g))|\Omega_\omega}\notag \\
    &\equiv -\frac{\partial^2}{\partial s\partial t}\Bigg|_{s,t=0}\!\!\!\!\!\!\!\!\braket{\Omega_\omega|e^{\ii s\pi_\omega(\hat\phi(f))}e^{\ii t\pi_\omega(\hat\phi(g))}|\Omega_\omega}\,.
\end{align}
In what follows we will thus write the formal two-point functions $\omega(\hat\phi(f)\hat\phi(g))$ with this understanding that the actual calculation is (implicitly) done with respect to the GNS representation in question.

\subsection{Quasifree states}

In the AQFT approach there are too many algebraic states available for us and not all of them are physically relevant. The general agreement among its practitioners is that all physically reasonable states associated to $\omega$ should be part of the class of \textit{Hadamard states} \cite{Khavkhine2015AQFT,KayWald1991theorems,Radzikowski1996microlocal}. Roughly speaking, these states have the right ``singular structure'' at short distances that respects the local flatness property in general relativity and that the expectation values of field observables are finite (see \cite{KayWald1991theorems,wald1994quantum} and references therein for more technical details). In this work, we would like to work with Hadamard states that are also \textit{quasifree}, denoted by $\omega_\mu$: these are the states which can be completely described only their two-point correlators\footnote{By this we mean that all odd-point functions vanish and only $\omega(\hat\phi(f)\hat\phi(g))\neq 0$. All even-point functions can be written as linear combination of products of two-point functions. The term \textit{Gaussian states} are sometimes reserved for states that have non-zero one-point correlators.}. Well-known field states such as the vacuum states and thermal states are all quasifree states, with thermal states (thermality defined according to Kubo-Martin-Schwinger (KMS) condition \cite{KayWald1991theorems}) being an example of mixed quasifree state.

The definition of quasifree states is somewhat tricky to work with, so we review it here following \cite{tjoa2022channel} (largely based on \cite{KayWald1991theorems,Khavkhine2015AQFT,fewster2019algebraic}). Any quasifree state $\omega_\mu$ is associated to a \textit{real inner product} $\mu:\Sol_\R(\M)\times \Sol_\R(\M)\to \R$ satisfying the inequality 
\begin{align}
    |\sigma(Ef,Eg)|^2\leq 4\mu(Ef,Ef)\mu(Eg,Eg)\,,
    \label{eq: real-inner-product}
\end{align}
for any $f,g\in \CS$. The state is pure if it saturates the above inequality appropriately \cite{wald1994quantum}. Then the quasifree state $\omega_\mu$ is defined as 
\begin{align}
    \omega_\mu(W(Ef)) \coloneqq e^{-\mu(Ef,Ef)/2}\,.
    \label{eq: quasifree}
\end{align}
We will drop the subscript $\mu$ and simply write $\omega$ in what follows. As stated, however, Eq. \eqref{eq: quasifree} is not helpful because it does not provide a way to calculate $\mu(Ef,Ef)$.

In order to obtain practical expression for the norm-squared $||Ef||^2 \coloneqq \mu(Ef,Ef)$, we first make the space of solutions of the Klein-Gordon equation into a Hilbert space\footnote{We will assume that the Hilbert space is already completed via its inner product.}. In \cite{KayWald1991theorems} it was shown that we can always construct a  \textit{one-particle structure} associated to quasifree state $\omega_\mu$, namely a pair $(K,\mathcal{H})$, where $\mathcal{H}$ is  a Hilbert space $(\mathcal{H},\braket{\cdot,\cdot}_\mathcal{H})$ together with an $\R$-linear map $K:\Sol_\R(\M)\to \mathcal{H}$ such that for $\phi_1,\phi_2\in \Sol_\R(\M)$
\begin{enumerate}[leftmargin=*,label=(\alph*)]
    \item $K\Sol_\R(\M)+\ii K\Sol_\R(\M)$ is dense in $\mathcal{H}$;
    \item $\mu(\phi_1,\phi_2)=\Re\braket{K\phi_1,K\phi_2}_\mathcal{H}$;
    \item $\sigma(\phi_1,\phi_2) = 2\Im\braket{K\phi_1,K\phi_2}_\mathcal{H}$.
\end{enumerate}
In the more usual language of canonical quantization, the linear map $K$ projects out the ``positive frequency part'' of real solution to the Klein-Gordon equation.
The smeared Wightman two-point function $\mathsf{W}(f,g)$ is then related to $\mu,\sigma$ by \cite{KayWald1991theorems,fewster2019algebraic}
\begin{align}
    \hspace{-0.05cm}\mathsf{W}(f,g) &\coloneqq \omega(\hat\phi(f)\hat\phi(g)) = \mu(Ef,Eg) + \frac{\ii}{2}E(f,g)\,,
\end{align}
where we have used the fact that $\sigma(Ef,Eg) = E(f,g)$.

Finally, by antisymmetry we have $E(f,f)=0$, hence
\begin{align}
    ||Ef||^2 =  \mathsf{W}(f,f) = \braket{KEf,KEf}_\mathcal{H}\,.
    \label{eq: algebraic-norm}
\end{align}
Therefore, we can compute $\mu(Ef,Ef)$ if either (i) we know the (unsmeared) Wightman two-point distribution of the theory associated to some quantum field state, or (ii) we know the inner product $\braket{\cdot,\cdot}_\mathcal{H}$ and how to project using $K$.

The inner product $\braket{\cdot,\cdot}_\mathcal{H}$ is precisely the \textit{Klein-Gordon inner product} $(\cdot,\cdot)_{\textsc{kg}}:\Sol_\C(\M)\times\Sol_\C(\M)\to \mathbb{C}$ restricted to $\mathcal{H}$, defined by
\begin{align}
    (\phi_1,\phi_2)_\textsc{kg} \coloneqq \ii \sigma(\phi_1^*,\phi_2)\,,
    \label{eq: KG-inner-product}
\end{align}
where the symplectic form is now extended to \textit{complexified} solution $\Sol_\C(\M)$ of the Klein-Gordon equation. The restriction to $\mathcal{H}$ is necessary since $(\cdot,\cdot)_\textsc{kg}$ is not an inner product on $\Sol_\C(\M)$. In particular, we have
\begin{align}
    \Sol_\C(\M) \cong \mathcal{H}\oplus \overline{\mathcal{H}}\,,
\end{align}
where $\overline{\mathcal{H}}$ is the complex conjugate Hilbert space of $\mathcal{H}$ \cite{wald1994quantum}. It follows that Eq.~\eqref{eq: quasifree} can be written as
\begin{align}
    \omega(W(Ef)) =   e^{-{\frac{1}{2}}\mathsf{W}(f,f)} = e^{-{\frac{1}{2}}||KEf||^2_{\textsc{kg}}}\,.
    \label{eq: norm-Ef}
\end{align}

The expression in Eq.~\eqref{eq: norm-Ef} gives us a concrete way to calculate $||Ef||^2$ more explicitly. For vacuum state, we know that the (unsmeared) Wightman two-point distribution is defined by
\begin{align}
    \mathsf{W}(\sx,\sy) &= \int \dd^3\bk\, u^{\phantom{*}}_\bk(\sx) u^*_\bk(\sy)\,,
\end{align}
where $u_\bk(\sx)$ are (positive-frequency) modes of Klein-Gordon operator $P$ normalized with respect to Klein-Gordon inner product Eq.~\eqref{eq: KG-inner-product}:
\begin{equation}
    \begin{aligned}
    (u_\bk,u_{\bk'})_\textsc{kg} &= \delta^3(\bk-\bk')\,,\quad (u_\bk^{\phantom{*}},u^*_{\bk'})_\textsc{kg} = 0\,,\\
    (u_\bk^*,u^*_{\bk'})_\textsc{kg} &= -\delta^3(\bk-\bk')\,.
    \end{aligned}
    \label{eq: KG-normalization}
\end{equation}
If we know the set $\{u_\bk\}$, we can  calculate the symmetrically smeared two-point function
\begin{align}
    \mathsf{W}(f,f) = \int \dd V\,\dd V' f(\sx)f(\sy)\mathsf{W}(\sx,\sy)\,.
    \label{eq: Wightman-double-smeared}
\end{align}
From the perspective of projection map $K$, what we are doing is projecting out the positive-frequency part of $Ef$ and express this in the positive-frequency basis $\{u_\bk\}$: that is, we have
\begin{align}
    Ef &= \int \dd^3\bk\, (u_\bk,Ef)_\textsc{kg} u_\bk + {(u_\bk,Ef)^*_\textsc{kg}u_\bk^*}\,,
\end{align}
so that using Eq.~\eqref{eq: KG-normalization} we get
\begin{align}
    KEf &= \int \dd^3\bk\, (u_\bk, Ef)_{\textsc{kg}} u_\bk(\sx)\,.
    \label{eq: KEf}
\end{align}
It follows that the restriction of the Klein-Gordon inner product to $\mathcal{H}$ gives
\begin{align}
    \braket{KEf,KEf}_\mathcal{H} &= (KEf,KEf)_\textsc{kg}\notag \\ 
    &= \int\dd^3\bk \,|(u_\bk,Ef)_\textsc{kg}|^2 \,.
    \label{eq: KG-norm-Ef}
\end{align}
Therefore, using the fact that \cite[Lemma 3.2.1]{wald1994quantum} ({See Appendix \ref{appendix: symplectic-smearing-Wald} for details})
\begin{align}
    \sigma(Ef,\phi) &= -\sigma(\phi,Ef) = \int\dd V f(\sx)\phi(\sx)\,, 
\end{align}
we can recast $(u_\bk,Ef)_\textsc{kg}$ as
\begin{align}
    (u_\bk , Ef)_\textsc{kg} &= \ii\sigma(u_\bk^*,Ef) = - \ii \int\dd V \, u_\bk^*(\sx)f(\sx)\,,
    \label{eq: basis-Ef}
\end{align}
so that indeed we recover $\braket{KEf,KEf}_\mathcal{H} = \mathsf{W}(f,f)$. 

The nice thing about algebraic formulation is that if we wish to consider another algebraic state, such as the thermal KMS state $\omega_\beta$ where $\beta$ labels the inverse KMS temperature, we will obtain a \textit{different} one-particle structure  $(K',\mathcal{H}')$. Hence the only thing that changes in the calculations so far is the replacement of $||Ef||^2$ in terms of the new one-particle structure. For thermal states, there is a nice expression for this in terms of the vacuum one-particle structure $(K,\mathcal{H})$ \cite{KayWald1991theorems}:
\begin{align}
    ||Ef||_{\beta}^2 
    &= \mathsf{W}_\beta(f,f)    
    \equiv \braket{K'Ef,K'Ef}_{\mathcal{H}'} \notag\\
    &= \braket{KEf,\coth(\beta\hat h/2)KEf}_{\mathcal{H}}\,,
\end{align}
where $W_\beta(f,f)$ is the smeared thermal Wightman distribution {(see, e.g., \cite[Chp. 2]{birrell1984quantum} for unsmeared version)} and $\hat h$ is the ``one-particle Hamiltonian'' (see also \cite{Landulfo2021cost}).

\section{Scalar QFT on $\skri^+$}
\label{sec: AQFT-null}

In this section our goal is to {review the construction of} scalar field quantization living on $\skri^+$. This necessarily requires us to restrict our attention to massless scalar fields since solutions to massive Klein-Gordon equation do not have support at $\skri$. Furthermore, we require that the field is conformally coupled to curvature in order to exploit good properties associated to Weyl rescaling of the bulk metric. Since we are working in $(3+1)$ dimensions, in what follows the real scalar field obeying Eq.~\eqref{eq: KGE} will be taken to have $m=0$ and $\xi = 1/6$. 

There are two reasons why scalar QFT on $\skri^+$ necessarily requires separate treatment. First, viewing $\skri^+$ as the conformal boundary of $
\M$, null infinity is a (codimension-1) null surface with degenerate metric (i.e., signature $(0,+,+)$). Second, the scalar QFT has \textit{no equation of motion} at $\skri^+$. Clarifying how this works is one of the main goals of this section. We will also connect how the AQFT framework relates to the more pedestrian (but perhaps more natural) approach used in {\textit{asymptotic quantization}, where one quantizes a bulk field theory and then perform near-$\skri^+$ expansion to obtain the corresponding boundary field theory}.

\subsection{Geometry of null infinity}

In order to set the stage, let us set up a few relevant definitions, in particular the notion of asymptotic flatness. 
We follow the rigorous definition in \cite{dappiaggi2015hadamard}
and explain what the conditions mean in practice \cite{Flanagan:2019vbl}.

{Let $(\M,g_{ab})$ be a globally hyperbolic manifold, which we call the \textit{physical spacetime}. We say that $(\M,g_{ab})$ is \textit{asymptotically flat with timelike infinity} $i^+$ if there exists an \textit{unphysical spacetime} $(\tilde{\M},\tilde{g}_{ab})$ with a preferred point $i^+\in \tilde\M$, a smooth embedding $F: \M \to \tilde\M$ (so that $\M$ can be viewed as embedded submanifold of $\tilde\M$), such that
\begin{enumerate}[label=(\alph*),leftmargin=*]
    \item The \textit{causal past} of $i^+$, denoted $J^-(i^+)$, is a closed subset of $\tilde\M$ such that 
    $\M = J^-(i^+)\setminus \partial J^-(i^+)$. The set $\skri^+\subset\tilde\M$ is called \textit{future null infinity} which is topologically $\R\times S^2$;
    
    \item There exists a smooth function $\Omega>0$ on $\tilde\M$, such that $\Omega|_{\skri^+}=0$, $\dd\Omega|_{\skri^+}\neq 0$, and 
    \begin{align}
        F^*(\Omega^{-2}\tilde{g}_{ab}) = g_{ab}\,,
        \label{eq: Weyl-rescaling}
    \end{align}
    typically written as $\tilde{g}_{ab} = \Omega^2 g_{ab}$. In the standard physics terminology, Eq.~\eqref{eq: Weyl-rescaling} is known as \textit{Weyl rescaling}\footnote{In \cite{wald2010general} it is called \textit{conformal transformation}, while angle-preserving diffeomorphism is called \textit{conformal isometry}. In high energy physics and AdS/CFT, conformal transformation often refers to angle-preserving diffeomorphism.}, typically written as $\tilde{g}_{ab} = \Omega^2 g_{ab}$, and $\Omega$ called the \textit{conformal factor} \cite{Flanagan:2019vbl}. At $i^+$, we have $\tilde\nabla_a\tilde\nabla_b\Omega = -2\tilde g_{ab}$ where $\tilde{\nabla}$ is the Levi-Civita connection with respect to the unphysical metric $\tilde{g}_{ab}$;

    \item Defining $n^a\coloneqq\tilde\nabla^a\Omega$, there exists a smooth positive function $\lambda$ supported at least in the neighbourhood of $\skri^+$ such that $\tilde\nabla_a(\lambda^4 n^a)|_{\skri^+} = 0$ and the integral curves of $\lambda^{-1}n^a$ are complete on $\skri^+$. 
    
    \item The physical stress-energy tensor $T_{ab}$ sourcing the Einstein field equation $G_{ab}=8\pi G T_{ab}$ obeys the condition $\tilde{T}_{ab} = \Omega^{-2} T_{ab}$,\footnote{{This condition formalizes the fact that {to be asymptotically flat the matter fields need to decay. For example,} even if we treat cosmological constant as the stress-energy tensor (i.e., $T_{ab} = -\Lambda g_{ab}$) instead of being a true cosmological constant, the spacetime is still not asymptotically flat.}} 
    where $\tilde{T}_{ab}$ is smooth on $\tilde{\M}$ and $\skri^+$. Note that for vacuum solutions this condition is redundant. 
\end{enumerate}

The four conditions mainly say the following: Condition (a) says that $\M$ lies in the causal past of its (future) boundary $\partial\M = \skri^+\cup \{i^+\}$, which is manifest when we draw Penrose diagrams; Condition (b) says that $\skri^+$ is the \textit{conformal boundary} of $\M$ and the conditions on $\Omega$ are technical ``price'' for bringing infinity into an actual boundary; Conditions (c) and (d) say that $\skri^+$ is a null hypersurface with normal $n^a$ and that Einstein equations are approximately vacuum at $\skri^+$ \cite{Flanagan:2019vbl}. Note that the technical condition $\tilde\nabla_a(\lambda^{4}n^a)=0$ is the statement that we can find null generators of $\skri^+$ that are divergence-free \cite{Ashtekar2018infraredissues}. This amounts to choosing the Bondi condition $\tilde{\nabla}_{a}n_b=0$ and implies $n_an^a= {\cal O}(\Omega^{-2})$~\cite{Flanagan:2019vbl}.
}

We can now state the properties of (future) null infinity $\skri^+$ that we are interested in \cite{Moretti2005BMS-invar,wald2010general}:
\begin{enumerate}[leftmargin=*,label=(\alph*)]
    \item {Since $\skri^+$ is a null hypersurface of $\tilde\M$ diffeomorphic to $\R\times S^2$, there exists an} open neighbourhood $U$ containing $\skri^+$ and a coordinate system $(\Omega,u,x^A)$ such that $x^A=(\theta,\varphi)$ defines standard coordinates of the unit two-sphere, $u$ is an affine parameter along the null geodesic of the generators of $\skri^+$. In this chart, $\skri^+$ is defined by the locus $\Omega=0$ and hence the metric reads
    \begin{align}
         \hspace{0.2cm} h\coloneqq g\bigr|_{\skri^+} &= (\dd\Omega\otimes \dd u+\dd u\otimes \dd\Omega) + \gamma_{S^2}\,,
        \label{eq: metric-null}
    \end{align}
    where $\gamma_{S^2}$ is the induced metric of $g$ on $S^2$, i.e.,
    \begin{align}
        \gamma_{S^2} &= \dd\theta\otimes\dd\theta+ \sin^2\theta\,\dd\varphi\otimes\dd\varphi\,.
    \end{align}
    The chart $(U,(\Omega,u,x^A))$ is called the \textit{Bondi chart}. 
    
    \item There exists a distinguished infinite-dimensional subgroup $\mathsf{BMS}_4(\skri^+)\subset \mathsf{Diff}(\skri^+)$, called the \textit{Bondi-Metzner-Sachs} (BMS) group, which leaves invariant the metric Eq.~\eqref{eq: metric-null}. This group is the semidirect product $SL(2,\C)\ltimes C^\infty(S^2)$. This is exactly the same group that preserves asymptotic symmetries of the physical spacetime $(\M,g_{ab})$ \cite{strominger2018lectures} (see Appendix~\ref{appendix: BMS} for more details).
\end{enumerate}

For completeness, we make a passing remark that this construction could have been generalized to other null surfaces, such as Killing horizons in black hole and cosmological spacetimes. The idea is to consider more generally the following ingredients \cite{Dappiaggi2009Unruhstate,Moretti2005BMS-invar,Dappiaggi2008cosmological}:
\begin{enumerate}[label=(\alph*),leftmargin=*]
    \item Let $\N\subset\tilde\M$ be a null hypersurface diffeomorphic to $\R\times \Xi$ with $\Xi$ a spacelike submanifold of $\M$. We can define the analogous Bondi chart $(\Omega,\lambda,x^A)$ on $\tilde\M$ so that on open neighbourhood $V$ containing $\N$, the hypersurface is the locus $\Omega=0$ so that $(\lambda,x^A)$ defines a coordinate system for $\N$. As before, we require $\dd\Omega|_{\N}\neq 0$.  
    
    \item The metric restricted to $\N$ takes analogous form to Eq.~\eqref{eq: metric-null}:
    \begin{align}
        h &= C^2 (\dd\Omega\otimes \dd \lambda+\dd \lambda\otimes \dd\Omega + \gamma_{\Xi})\,,
        \label{eq: metric-null-2}
    \end{align}
    where $C\neq 0$ is real. As before $\lambda$ will define an affine parameter for null generators of $\N$.  
\end{enumerate}
In this sense, the structure of null infinity and Killing horizons are very similar. For example, the \textit{future horizon} $\hor^+$ of Schwarzschild geometry is associated to $\lambda = U$ where $U$ is one of the the Kruskal-Szekeres coordinates, with $C\neq 1$ (unlike the case for $\skri^+$). There is some extra care that one needs to be aware of for metrics that contain horizons, but in this work we will not consider these cases and leave it for future investigations. We direct interested readers regarding the same constructions involving horizons to \cite{Dappiaggi2009Unruhstate,Moretti2005BMS-invar}.

\subsection{Quantization at null infinity}

Next we try to construct scalar field theory at $\skri^+$. The main subtlety compared to standard bulk scalar theory is that $\skri^+$ is a null submanifold with degenerate metric, and that we should consider the equivalence classes of the triple $[(\skri^+,h,n)]$, where $(\skri^+,h,n)\sim(\skri^+,h',n')$ if they are related by a transformation in $\mathsf{BMS}_4(\skri^+)$. This latter condition is the statement that the $\mathsf{BMS}_4(\skri^+)$ is an asymptotic symmetry of all asymptotically flat spacetimes and {$\skri^+$ is a universal structure of these spacetimes} \cite{Flanagan:2019vbl} (see Appendix~\ref{appendix: BMS}). For these reasons, the scalar field theory at null infinity will ``look'' different from the bulk theory, but procedurally the construction proceeds the same way, as we will show.

First, fix a Bondi frame $(\skri^+,h,n)$. We define a real vector space of ``solutions''\footnote{Although there is no equation of motion at $\skri^+$, we denote the real vector space $\Sol_\R(\skri^+)$ this way because as we will see it is related to the space of solutions in the bulk.} \cite{dappiaggi2015hadamard} 
\begin{align}
    \Sol_\R(\skri^+)\coloneqq \{\psi\in C^\infty(\skri^+):\psi,\partial_u\psi\in L^2(\skri^+,\dd\mu)\}\,.
    \label{eq: skri-sol}
\end{align}
where $\dd\mu =\dd u\; {\dd \gamma_{S^2}}$ is {the integration measure}, ${\dd \gamma_{S^2}}$  the standard volume form on $S^2$, and $L^2(\skri^+,\dd\mu)$ is the space of square-integrable functions with respect to 
$\dd\mu$. This space becomes a symplectic vector space if we give it a symplectic form $ \sigma_\skri:\Sol_\R(\skri^+)\times\Sol_\R(\skri^+)\to \R$ with
\begin{align}\label{eq: skri-symplc}
   \sigma_\skri(\psi_1,\psi_2)=\int_\skri \dd\mu\, (\psi_{{1}}\partial_u \psi_{{2}}-\psi_{{2}}\partial_u \psi_{{1}})\,. 
\end{align}
The symplectic structure is independent of the choice of Bondi frames \cite{Moretti2005BMS-invar} (we reproduce the essential features to demonstrate this in Appendix \ref{appendix: BMS}). We can then define a ``Klein-Gordon'' inner product
\begin{align}
    (\psi_1,\psi_2)_{\skri} \coloneqq  \ii\sigma_\skri(\psi_1^*,\psi_2)\,.
    \label{eq: KG-like-inner-product}
\end{align}

Recall from Section~\ref{sec: AQFT} that in order to obtain the quantization for the bulk scalar theory, we needed the algebra of observables $\A(\M)$ (or $\W(\M)$) and an algebraic state $\omega$. For quasifree states $\omega_\mu$ defined by a real symmetric bilinear inner product $\mu$ on $\Sol_\R(\M)$, the characterization of $\omega_\mu$ depends on the one-particle structure $(K,\mathcal{H})$. The Hilbert $\mathcal{H}$ is essentially the ``positive frequency subspace'' of the \textit{complexified} solution space $\Sol_\C(\M)$, with inner product given by Klein-Gordon inner product extended to the complex domain. As we will see now, the definition of $(\Sol_\R(\skri^+),\sigma_\skri)$ essentially lets us carry the same procedure almost verbatim.

\subsection{Algebra of observables}

Similar to the bulk algebra of observables, we have the \textit{boundary algebra of observables} $\A(\skri^+)$ whose elements are generated by unit $\openone$ and the \textit{smeared boundary field operator} $\varphi(\psi)$, where $\psi\in C_0^\infty(\skri^+)$\,. However, there are several structural differences. First, there is no equation of motion at $\skri^+$, so $\A(\skri^+)$ is defined differently from $\A(\M)$. Second, the metric is degenerate at $\skri^+$ and hence the smeared field operator $\varphi(\psi)$ has to be defined carefully.

That said, we can still work directly with the Weyl algebra corresponding to ``exponentiated'' version of $\A(\skri^+)$, denoted $\W(\skri^+)$, where many things are better behaved. This is because given a symplectic vector space $(\Sol_\R(\skri^+),\sigma_\skri)$, there exists a complex $C^*$-algebra generated by elements of $\Sol_\R(\skri^+)$ \cite{bratteli2002operatorv1}. The Weyl algebra $\W(\skri^+)$ is generated by $\openone$ and $W(\psi)$ for $\psi\in \Sol_\R(\skri^+)$. The Weyl relations are\footnote{We will not distinguish the notation of the elements of the Weyl algebra in the bulk and in the boundary and use $\W(\cdot)$ for the Weyl algebra and $W(\cdot)$ as its elements. The bulk elements will always be written as $W(Ef)$ with causal propagator $E$, while the boundary element will be written as $W(\psi)$.}
\begin{subequations}
\begin{align}
    W(\psi)^\dagger &= W(-\psi)\,, \\ 
    W(\psi)W(\psi') &= {e^{-\ii\sigma_\skri(\psi,\psi')/2}}W(\psi+\psi')\,.
\end{align}
\label{eq: Weyl-relations-skri}
\end{subequations}
This Weyl algebra is unique up to (isometric) $*$-isomorphism \cite{bratteli2002operatorv1}. Note that $\W(\skri^+)$ contains unit element associated to $\psi=0$, and $W(\psi)$ is uniquely specified by $\psi$. Moreover, since there is no equation of motion on $\skri^+$, there is no causal propagator and hence the locality condition (often called \textit{microcausality} in QFT) is not implemented by the causal propagator; instead, this can be imposed using definition of $\sigma_\skri$ by
\begin{align}
    [W(\psi),W(\psi')] = 0\quad \supp(\psi)\cap\supp(\psi') = \emptyset\,.
\end{align}
Note that this is exactly the same locality relation as in the bulk theory since there we have $\sigma(Ef,Eg) = E(f,g)$ 
(\textit{c.f.} Eq.~\eqref{eq: Weyl-relations}).

\subsection{Quasifree state at $\skri^+$}

Now, let us construct a one-particle structure for $(\sK,\mathfrak{H})$ for $\Sol_\R(\skri^+)$. We will follow closely the construction in \cite{Moretti2005BMS-invar}, focusing on accessibility for physics-oriented readers. 

First, define $\sK:\Sol_\R(\skri^+) \to \mathfrak{H}$ to be the positive-frequency projector given by
\begin{align}
    (\sK \psi)(u,x^A) &= \frac{1}{\sqrt{2\pi}}\int_0^\infty\!\!\!\dd \omega\,e^{-\ii\omega u}\tilde{\psi}(\omega,x^A)\,, \\
    \tilde{\psi}(\omega,x^A) &= \frac{1}{\sqrt{2\pi}}\int_{-\infty}^\infty\!\!\!\dd u\,e^{\ii\omega u}\tilde{\psi}(u,x^A)\,.
\end{align}
That is, $\tilde\psi$ is the $u$-domain Fourier transform of $\psi$ {and hence $\sK\Sol_\R(\skri^+) + \ii \sK\Sol_\R(\skri^+)$ is dense in $\Sol_\C(\skri^+)$}. The space $\mathfrak{H}$ is a Hilbert space with inner product defined as restriction to ``Klein-Gordon'' inner product Eq.~\eqref{eq: KG-like-inner-product}
\begin{align}
    \braket{\sK\psi_1,\sK\psi_2}_\mathfrak{H} \coloneqq  (\sK\psi_1,\sK\psi_2)_\skri\,.
\end{align}
It was shown in \cite{Moretti2005BMS-invar} that there exists a \textit{ $\mathsf{BMS}_4(\skri^+)$-invariant quasifree and regular algebraic state} (see also \cite{Khavkhine2015AQFT} for definition of regular state) $\omega_\skri:\W(\skri^+)\to \C$ such that
\begin{align}
    \omega_\skri(W(\psi)) = e^{-\mu_\skri(\psi,\psi)/2}\,,
\end{align}
where $\mu_\skri:\Sol_\R(\skri^+)\times \Sol_\R(\skri^+)\to \R$ is a real bilinear inner product given by
\begin{align}
    \mu_\skri(\psi_1,\psi_2) = \Re\braket{\sK\psi_1,\sK\psi_1}_\mathfrak{H}\,.
\end{align}
Notice that up to this point, the procedure exactly parallels that of the bulk scalar field theory.

In fact, we can be very explicit about this algebraic state. First, since we already have the algebraic state $\omega_\skri$ and the Weyl algebra of observables $\W(\skri^+)$, we can use GNS theorem to construct the Fock representation of the boundary field. Recalling that in the GNS representation we can take derivatives of the representation of the Weyl algebra (\textit{c.f.} Eq.~\eqref{eq: Wightman-formal-bulk}), we can calculate the \textit{smeared Wightman two-point function} at $\skri^+$ \cite{dappiaggi2015hadamard}:
\begin{align}
    &\mathsf{W}_\skri(\psi_1,\psi_2)
    {{\,\,``\!\!\coloneqq\!\!"\,\, \omega_\skri\left(\hat\varphi(\psi_1)\hat\varphi(\psi_2)\right)}} \notag\\
    &= -\frac{1}{\pi}\lim_{\epsilon\to 0}\int {\dd \gamma_{S^2}}\dd u\dd u'\frac{\psi_1(u,x^A)\psi_2(u',x^A)}{(u-u'-\ii\epsilon)^2}\,.
    \label{eq: boundary-Wightman}
\end{align}
{The definition of $\omega_\skri\left(\hat\varphi(\psi_1)\hat\varphi(\psi_2)\right)$ requires that we define what ``boundary field'' $\varphi(\psi)$ with boundary smearing function $\psi$ means. We will clarify this point in Section~\ref{sec: large-r-expansion}.} Note in particular that the integral is taken over the same angular direction 
$x^A$ for $\psi_1$ and $\psi_2$. Eq.~\eqref{eq: boundary-Wightman} is the main result we will use for our holographic reconstruction.

\subsection{Modest holography: bulk-to-boundary correspondence}

At this point, the Weyl algebras $\W(\M)$ and $\W(\skri^+)$ as well as the space of solutions $\Sol_\R(\M)$ and $\Sol_\R(\skri^+)$ are unrelated, so the two scalar field theories are \textit{a priori} unrelated. Indeed, it is not automatic that one can establish some sort of holographic principle or bulk-to-boundary correspondence between them. The reason is because for this to work, we need to ``project'' bulk solutions to $\skri^+$, i.e., we need the \textit{existence} of a projection map $\Gamma:\Sol_\R(\M)\to \Sol_\R(\skri^+)$. This is necessary in order for an injective $*$-homomorphism $i:\W(\M)\to \W(\skri^+)$ to exist and build the bulk-to-boundary correspondence.

The celebrated result in \cite{Dappiaggi2005rigorous-holo} shows that the boundary Weyl algebra is in fact very natural: this is because one can prove that \textit{if} there exists a projection map $\Gamma:\Sol_\R(\M)\to\Sol_\R(\skri^+)$ such that
\begin{enumerate}[leftmargin=*,label=(\arabic*)]
    \item The bulk solutions projected to $\skri^+$ lies in $\Sol_\R(\skri^+)$, i.e., $\Gamma\Sol_\R(\M)\subset \Sol_\R(\skri^+)$;
    \item The symplectic forms are compatible with the projection, i.e., {$\sigma(\phi_1,\phi_2) = \sigma_\skri(\Gamma\phi_1,\Gamma\phi_2)$},
\end{enumerate}
then the bulk algebra $\W(\M)$ can be identified with a $C^*$-subalgebra of $\W(\skri^+)$, in that there exists an isometric $*$-isomorphism $\iota:\W(\M)\to \iota(\W(\M))\subset \W(\skri^+)$ such that
\begin{align}
    \iota(W(Ef)) = W(\Gamma Ef) \in \W(\skri^+)\,.
    \label{eq: injective-star}
\end{align}
Furthermore, $(\Sol_\R(\skri^+),\sigma_\skri)$ is \textit{universal} for all asymptotically flat spacetimes $\M$. These conditions guarantee that the Weyl algebras are compatible, i.e., the bulk scalar field in $\M$ can be ``holographically'' projected to the boundary $\skri^+$ and defines a boundary scalar field there. 

The injective $*$-homomorphism $\iota:\W(\M)\to \W(\skri^+)$ can be used to perform pullback on the algebraic state $\omega_\skri$. That is, the state $ \omega\coloneqq (\iota^*\omega_\skri):\W(\M)\to\C$ 
is an algebraic state on $\W(\M)$, with the property \cite{dappiaggi2012rigorous}
\begin{align}
    \omega(W(Ef))\equiv (\iota^*\omega_\skri)(W(Ef)) = \omega_\skri(W(\Gamma Ef))\,,
    \label{eq: pullback-state}
\end{align}
in accordance to Eq.~\eqref{eq: injective-star}. This result is remarkable because (i) the $\mathsf{BMS}_4(\skri^+)$-invariant state $\omega_\skri$ is \textit{unique} (in its folium), thus the algebraic state $\iota^*\omega_\skri$ is also unique~\cite{Moretti2005BMS-invar}; (ii) the pullback state $\omega$ is Hadamard and is invariant under \textit{all} isometries of $(\M,g_{ab})$~\cite{Moretti2008outstates}. In the case when the bulk geometry is flat, $\iota^*\omega_\skri$ would define what we know as Poincar\'e-invariant Minkowski vacuum.

It is worth stressing that this construction relies on the {existence} of the projection map $\Gamma$ whose image lives entirely in $\Sol_\R(\skri^+)$. This assumption is not automatic, and we can think of three representative examples:
\begin{enumerate}[align=left,leftmargin=*, label=(\roman*)]
    \item In Schwarzschild spacetime, we also have Killing horizons $\mathscr{H}^\pm$, thus $\skri^+$ alone is not enough, we also need $\mathscr{H}^+$ to build the correspondence \cite{Dappiaggi2009Unruhstate}. 
    
    \item In Friedmann-Robertson-Walker (FRW) spacetimes, we also have cosmological horizons $\mathscr{H}^\pm_{\text{cosmo}}$ which play the role of null infinity $\skri^\pm$ even if the spacetime is not asymptotically flat \cite{Dappiaggi2008cosmological}. Since the geometry is asymptotically de Sitter, it is impossible to build the correspondence this way for the entire de Sitter hyperboloid. For matter- or radiation-dominated FRW models which are asymptotically flat, it is still possible that some information is lost into the timelike infinity $i^\pm$.

    \item In spacetimes containing ergoregions\footnote{We thank Gerardo Garc\'ia-Moreno for pointing this out.}, it is possible for some bulk solutions to get projected into future timelike infinity $i^+$ instead of $\skri^+$, essentially due to the asymptotic time-translation Killing field becoming spacelike. 
\end{enumerate}
In all these cases, the key observation is that it is not automatic that if $Ef\in \Sol_\R(\M)$ then it can be projected properly to $\Sol_\R(\skri^+)$: more concretely, in terms of Bondi coordinates, the ``conformally rescaled'' boundary data
\begin{align}
    \psi_f \coloneqq \lim_{\substack{r\to\infty\\
    u\text{ const.}}} \Omega^{-1}Ef
\end{align}
may not be an element of $\Sol_\R(\skri^+)$. It is in this sense that in general null infinity is not a good initial data surface \cite{Geroch:1978us}. Even for globally hyperbolic spacetimes without horizons, one typically needs to augment $\skri^+$ with future timelike infinity $i^+$ to make this work {(also see \cite{prabhu2022infrared} and references therein). In what follows we will work with the assumption that the spacetime is one where the bulk-to-boundary correspondence \eqref{eq: injective-star} holds.}

\section{Holographic reconstruction of the bulk metric}
\label{sec: holographic-reconstruction}

The injective $*$-homomorphism $\iota$ allowed us to define the bulk algebraic state $\omega$ via the pullback of algebraic state $\omega_\skri$ in Eq.~\eqref{eq: pullback-state}. We also know that the elements of the Weyl algebra are formally the ``exponentiated'' version of the smeared field operator $\phi(f)$. Therefore, in order for us to say that we can perform holographic reconstruction, we require that for $f,g\in\CS$, we can construct the smeared Wightman function in the bulk in the sense given in Section~\ref{sec: AQFT} such that it agrees with the boundary via the relation:
\begin{align}
    \mathsf{W}(f,g) = \mathsf{W}_\skri(\psi_f,\psi_g)\,,
    \label{eq: wightman-holographic-reconstruction}
\end{align}
where $\psi_f = \Gamma Ef$ and $\psi_g = \Gamma Eg$. 

The holographic reconstruction is complete once we modify the result from  Saravani, Aslanbeigi and Kempf \cite{Kempf2016curvature,Kempf2021replace} to reconstruct the metric from $\mathsf{W}(f,g)$ instead of the Feynman propagator (or equivalently, following the analogous proposal  in \cite{perche2021geometry})\footnote{There the focus was on measurement of of metric components using Unruh-DeWitt detectors which can probe the correlators.}. The idea  is that one can reconstruct the metric formally by computing in (3+1) dimensions the ``coincidence limit'' of the inverse Wightman function
\begin{align}
    g_{\mu\nu}(\sx) = -\frac{1}{8\pi^2}\lim_{\sx' \to \sx}\partial_\mu\partial_{\nu'} \mathsf{W}(\sx,\sx')^{-1}\,.
    \label{eq: metric-from-wightman}
\end{align}
Notice that this is not surprising (in hindsight!) because causal propagators and Green's functions know about the metric function directly. In particular, in the case of Wightman functions, the requirement that the states are Hadamard means that for closely separated events $\sx,\sy$ the Wightman function is of the form \cite{wald1994quantum}
\begin{align}
    &\mathsf{W}_{\M}(\sx,\sy) \notag\\
    &= \frac{U(\sx,\sy)}{8\pi^2\sigma_\epsilon(\sx,\sy)}+V(\sx,\sy)\log\sigma_\epsilon(\sx,\sy)+Z(\sx,\sy)\,,
    \label{eq: Hadamard-form}
\end{align}
where $U,V,Z$ are regular smooth functions and $U\to1$ as  $\sx \to \sy$. The bi-scalar $\sigma_\epsilon(\sx,\sy)$ is the Synge world function with $i\epsilon$ prescription, i.e.,
\begin{subequations}
\begin{align}
    \sigma(\sx,\sy) &= \frac{1}{2}(\tau_\sy-\tau_\sx)\int_\gamma g_{\mu\nu}(\lambda)\dot{\gamma}^\mu(\lambda)\dot{\gamma}^\nu(\lambda)\dd\lambda\,,\\
    \sigma_\epsilon(\sx,\sy) &= \sigma(\sx,\sy) +2i\epsilon(T(\sx)-T(\sy)) +\epsilon^2\,,
\end{align}
\end{subequations}
where $\sigma(\sx,\sy)\equiv \sigma_{\epsilon=0}(\sx,\sy)$ is the Synge world function, $T$ is a global time function (which exists by virtue of global hyperbolicity of $\M$) and $\gamma(\tau)$ is a geodesic curve with affine parameter $\tau$ with 
$\gamma(\tau_\sx)=\sx$ and $\gamma(\tau_\sy)=\sy$. Schematically, Eq.~\eqref{eq: metric-from-wightman} comes from the fact that when $\sy\approx \sx$ we have $\Delta\sx=\sx-\sy\approx 0$ and 
\begin{align}
    \mathsf{W}_\M(\sx,\sy)^{-1} &\approx {8\pi^2}\sigma(\sx,\sy) \notag\\
    &\sim {4\pi^2} g_{\mu\nu}(\sx) \Delta x^\mu \Delta x^\nu + \mathcal{O}(\Delta\sx^2)\,.
\end{align}

Our calculations in the previous sections treat the Wightman two-point functions as \textit{smeared} two-point functions. In practice this means that the expression in Eq.~\eqref{eq: metric-from-wightman} should be computed as difference equation centred around the peak of the smearing functions. Furthermore, the smearing implies that there is a ``resolution limit'' directly defined by the supports of the smearing functions $f,g$. Physically we can interpret this as the statement that vacuum noise prevents us from reconstructing the metric with infinite accuracy. Taking this into account, we calculate the metric using finite difference: let $f,g$ to be sharply peaked functions with characteristic widths $\delta$ localized around $\sx$ and $\sy$ respectively\footnote{We can take $f,g$ to be Gaussian as an approximation since the tails quickly become negligible and are effectively compactly supported and $\delta$ measures the width of the Gaussian. This allows for more controlled calculations in what follows.}. The finite-difference approximation of $\partial_\mu\partial_{\nu'}\mathsf{W}(\sx,\sx')^{-1}$ applied to the reciprocal of the Wightman function reads 
\begin{widetext}
\begin{align}
    \partial_\mu\partial_{\nu'}\mathsf{W}(\sx,\sx')^{-1}
    &\approx \frac{\mathsf{W}(\sx+\epsilon^\mu,\sx'+\epsilon^{\nu
    '})^{-1}-\mathsf{W}(\sx+\epsilon^\mu,\sx')^{-1}}{\delta^2}-\frac{\mathsf{W}(\sx,\sx'+\epsilon^{\nu'})^{-1}-\mathsf{W}(\sx,\sx')^{-1}}{\delta^2}\,.
    \label{eq: finite-difference-scheme}
\end{align}
\end{widetext}
Here the vector $\epsilon^\mu$ points in the direction of coordinate basis $\partial_\mu$ with very small length $\sqrt{|\epsilon^\mu\epsilon_\mu|} = \delta \ll 1$. A change of variable (shift by $\epsilon^\mu$) and smearing the Wightman functions before taking its reciprocal allows us to write the metric approximation as
\begin{align}
    g_{\mu\nu}(\sx) &\approx {-\frac{1}{8\pi^2\delta^2}}\Bigr[\mathsf{W}(f_\epsilon,g_\epsilon)^{-1} - \mathsf{W}(f_\epsilon,g)^{-1}
    \notag\\
    &\hspace{1.4cm}-\mathsf{W}(f,g_\epsilon)^{-1}+\mathsf{W}(f,g)^{-1}\Bigr]\,,
    \label{eq: metric-reconstruction}
\end{align}
where $f_\epsilon(\sx)=f(\sx-\epsilon^\mu)$ and $g_\epsilon(\sx) = g(\sx-\epsilon^{\nu'})$. The approximation improves with smaller $\delta$ but this is bounded below by the resolution provided by characteristic widths of $f,g$. Also note that the spacetime smearing functions must be properly normalized to reproduce the metric.

For our purposes, however, we want to make this reconstruction work from the boundary. So what we would like to calculate is $\mathsf{W}_\skri(\psi_1,\psi_2)$ in Eq.~\eqref{eq: boundary-Wightman}, use that to reconstruct $\mathsf{W}(f,g)$ using bulk-to-boundary correspondence \eqref{eq: wightman-holographic-reconstruction}, and then reconstruct the metric by finite difference scheme \eqref{eq: finite-difference-scheme}. From Eq.~\eqref{eq: boundary-Wightman}, we see that what really remains to be done is to compute $\Gamma : \Sol_\R(\M)\to\Sol_\R(\skri^+)$. 
When this projection map exists, its action is quite simple in Bondi chart: it is given by\footnote{In standard language of asymptotic symmetries literature, $\psi_f$ constitutes a \textit{boundary data} for the bulk scalar field theory \cite{strominger2018lectures}.} 
\begin{align}
    \psi_f\equiv (\psi_f)(u,x^A) = \lim_{r\to\infty}  (\Omega^{-1}Ef)(u,r,x^A)\,,
    \label{eq: boundary-data}
\end{align}
where $\Omega = 1/r$.

The final step to obtain the holographic reconstruction is to combine modest holography with the metric reconstruction using bulk correlators
---crucially, the state induced in the bulk by \eqref{eq: pullback-state} has Hadamard property \cite{Moretti2005BMS-invar}. That is, using Eq.~\eqref{eq: pullback-state} and \eqref{eq: metric-reconstruction} we obtain
\begin{align}
    g_{\mu\nu}(\sx) &\approx -{\frac{1}{8\pi^2\delta^2}}\Bigr[\mathsf{W}_\skri(\psi_{f_\epsilon},\psi_{g_\epsilon})^{-1}- \mathsf{W}_\skri(\psi_{f_\epsilon},\psi_{g})^{-1} \notag\\
    &\hspace{0.4cm}-\mathsf{W}_\skri(\psi_{f},\psi_{g_\epsilon})^{-1}+\mathsf{W}_\skri(\psi_{f},\psi_{g})^{-1}\Bigr]\,,
    \label{eq: metric-reconstruction-boundary}
\end{align}
where as before $\psi_{f_\epsilon}=\Gamma (Ef_{\epsilon})$. Eq.~\eqref{eq: metric-reconstruction-boundary} tells us how to reconstruct the bulk metric from the boundary Wightman function of the scalar field at $\skri^+$. This is the main result of this paper. 

On a more practical issue, the `bottleneck' of the holographic reconstruction is the \textit{classical} component, namely the causal propagator $E$: the holographic reconstruction is as simple or as hard as the computability of the causal propagator and its action on compactly supported test function $f$. Furthermore, it also relies on hardness of computing the integral, which in turn depends on the boundary smearing functions.

In this work we will restrict our attention computing the bulk and boundary correlators for two simple examples which are transparent and manageable yet physically relevant: (1) Minkowski space, and (2) Friedmann-Robertson-Walker (FRW) universe conformally related to Minkowski space. For Minkowski space, we will show how the bulk metric can be reconstructed from its boundary explicitly, since there is exact closed-form expression for the bulk/boundary correlator\footnote{This is much harder task than recovering the bulk metric from its bulk \textit{unsmeared} correlator, as done in \cite{perche2021geometry,Kempf2016curvature}, which can be done quite easily, as we will see later. {The problem is that the \textit{unsmeared} boundary correlator is ``universal'' (see Section~\ref{sec: large-r-expansion})}.}. For the FRW case, we will content ourselves with showing that the bulk-to-boundary reconstruction works by showing that the boundary and bulk correlators agree since the remaining obstruction is merely numerical in nature.

\subsection{Example 1: Minkowski spacetime}

\begin{figure}[tp]
    \centering
    \includegraphics[scale=1]{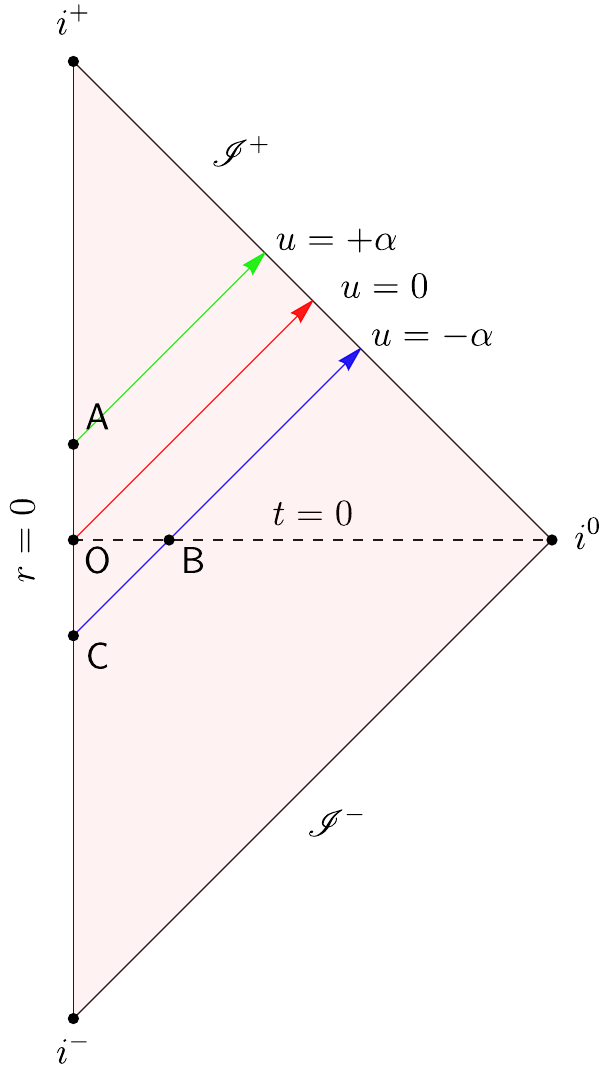}
    \caption{Penrose diagram for the bulk-to-boundary reconstruction in Minkowski space.}
    \label{fig: penrose-reconstruction}
\end{figure}

For the metric reconstruction, it is useful to first give the Penrose diagram as shown in Figure~\ref{fig: penrose-reconstruction}. Due to spherical symmetry, we can consider the holographic reconstruction to work if we can reconstruct the bulk Wightman function for three types of pair of events: one for timelike pairs, one for null pairs, and one for spacelike pairs. For convenience, let us fix the following four points in Bondi coordinates $(t,r,x^A)$, setting $x^A=0$ by spherical symmetry:
\begin{equation}
    \begin{aligned}
    \textsf{O} &= (0,0,0,0)\,,\quad \textsf{A} = (\alpha,0,0,0)\,,\\
    \textsf{B} &= (0,\alpha,0,0)\,,\quad \textsf{C} = (-\alpha,0,0,0)\,. 
    \end{aligned}
\end{equation}
Without loss of generality we can consider the timelike pair to be $\textsf{OA}$, the spacelike pair to be $\textsf{OB}$, and the null pair to be $\textsf{BC}$ (essentially due to translational and rotational invariance).

For simplicity, let us consider four distinct spacetime smearing functions
\begin{align}
    f_j\equiv f_j(\sx) = \chi\rr{\frac{t-t_j}{T_j}}\delta^3(\bx-\bx_j)\,,
\end{align}
where $\chi(\tau)$ is chosen to be a smooth function with a peak centred at $\tau=0$, $j= \textsf{O,A,B,C}$ labels the points in the bulk geometry for which $f_j$ is localized, $T_j$ labels the characteristic timescale of interaction. Thus the spacetime smearing $f_j$ is very localized in space and slightly smeared in time. For concrete calculations, let us fix the switching function to be a \textit{normalized} Gaussian, so that\footnote{Note that the Gaussian switching renders $\supp(f_j) \not\in\CS$, given any open neighbourhood $O_j$ of $\Sigma_{t_j}$ we can always choose $T$ small enough so that $\supp(f_j)$ centred at $t=t_j$ and $\bx=\bx_j$ is for all practical purposes compactly supported in $O_j$.}
\begin{align}
    \chi_j(t)\coloneqq \chi\rr{\frac{t-t_j}{T_j}} = 
    \frac{1}{\sqrt{\pi T_j^2}} e^{-(t-t_j)^2/T_j^2}\,,
\end{align}
and for simplicity we set $T_j=T$ for all $j$. For the time being we set $\lambda=1$.  In flat space, this choice enables us to compute the smeared Wightman function in closed form:
\begin{align}
    &\mathsf{W}(f_i,f_j) = \frac{1}{\sqrt{128 \pi^3 T^2} \left|\Delta\bx_{ij}\right|}e^{-\tfrac{\left|\Delta\bx_{ij}\right|^2+\left(\Delta t_{ij}\right)^2}{T^2}}\notag\\
    &\times
    \Bigg[e^{\tfrac{\left(\left|\Delta\bx_{ij}\right| + \Delta t_{ij}\right)^2}{2 T^2}} \left(\text{erfi}\left[\frac{\left|\Delta\bx_{ij}\right| - \Delta t_{ij}}{\sqrt{2} T}\right]+i\right)\notag\\
    &\hspace{0.2cm} +e^{\tfrac{\left(\left|\Delta\bx_{ij}\right| -\Delta t_{ij}\right){}^2}{2 T^2}} \left(\text{erfi}\left[\frac{\left|\Delta\bx_{ij}\right| + \Delta t_{ij}}{\sqrt{2} T}\right]-i\right)\Bigg]\,,
    \label{eq: bulk-VEV-minkowski-gauss}
\end{align}
where $\Delta t_{ij} = t_j - t_i$ and $\Delta \bx_{ij} = |\bx_j-\bx_i|$.

In order to calculate the boundary Wightman function, we need the causal propagator. The causal propagator in flat space is given by
\begin{align}
    E(\sx,\sy) 
    &= \frac{\delta(\Delta t+|\Delta\bx|) - \delta(\Delta t-|\Delta \bx|)}{4\pi|\Delta\bx|}
    \label{eq: causal-propagator-flat}
\end{align}
where $\Delta t = t-t'$ and $\Delta\bx = |\bx-\by|$. Using the modified null coordinates \eqref{eq: modified-null-coord}, we get
\begin{align}
    Ef_j(\sx) &= \int \dd^4\sy\,E(\sx,\sy)f_j(\sy)\notag\\
    &= \frac{\chi\rr{\tfrac{t-t_j+|\bx-\bx_j|}{T}} - \chi\rr{\tfrac{t-t_j-|\bx-\bx_j|}{T}}}{4\pi|\bx-\bx_j|}\,.
\end{align}
We can introduce a ``modified null variables'' $u_j,v_j$ defined by
\begin{equation}
    \begin{aligned}
    u_j &\coloneqq t-t_j-|\bx-\bx_j|\,,\\
    v_j &\coloneqq t-t_j+|\bx-\bx_j|\,,
    \label{eq: modified-null-coord}
    \end{aligned}
\end{equation}
so that $Ef_j$ takes a simple form
\begin{align}
    Ef_j(\sx)
    &= \frac{1}{4\pi|\bx-\bx_j|}\Bigr[\chi\rr{\frac{v_j}{T}} - \chi\rr{\frac{u_j}{T}}\Bigr]\,.
    \label{eq: Efj-bulk}
\end{align}
The boundary data associated to $Ef_j$, denoted by $\varphi_j$, is is the projection of $Ef_j$ to $\skri^+$ via the projection map $\Gamma$. This is done by taking the limit $r=|\bx| \to \infty$ while fixing $u=t-r$ constant (or $v=t+r \to \infty$ while fixing $u$ constant in double-null coordinates), so that
\begin{align}
    \Gamma Ef = \lim_{r\to\infty} \Omega^{-1}Ef\,,\quad\Omega = \frac{1}{r}\,.
\end{align}
In this limit, the modified null variables become
\begin{equation}
    \begin{aligned}
    u_j &\to u-(t_j - |\bx_j|\cos\theta_j) \,,\\
    v_j &\to v-(t_j + |\bx_j|\cos\theta_j)\,,
    \label{eq: modified-null-coord-boundary}
    \end{aligned}
\end{equation}
where $\theta_j$ is the angle between $\bx$ and $\bx_j$. 

The modest holography amounts to the claim that $\mathsf{W}_\skri(\varphi_{i},\varphi_{j})$ for Gaussian smearing is also given by Eq.~\eqref{eq: bulk-VEV-minkowski-gauss}. Let us see how this works concretely using examples. For brevity we will compute just one timelike pair and one spacelike pair explicitly, and one can check that it will work in general.

\subsubsection{Timelike pair \emph{$\textsf{OA}$}}

For point $\textsc{O}$, we have $s_\textsf{O}=0$ and $\bx_\textsf{O}=\mathbf{0}$, hence
\begin{align}
    Ef_\textsf{O} = \frac{1}{4\pi r}\Bigr[\chi\rr{\frac{v}{T}}-\chi\rr{\frac{u}{T}}\Bigr]\,.
\end{align}
It follows that the boundary data is
\begin{align}
    \varphi_\textsf{O} =  -\frac{1}{4\pi}\chi\rr{\frac{u}{T}}\,.
\end{align}
For point $\textsf{A}$, we have 
\begin{align}
    Ef_\textsf{A} = \frac{1}{4\pi r}\Bigg[\chi\rr{\frac{v-\alpha}{T}}-\chi\rr{\frac{u-\alpha}{T}}\Bigg]\,,
\end{align}
The boundary data associated to $Ef_\textsc{O}$ reads
\begin{align}
    \varphi_\textsf{A} =  -\frac{1}{4\pi}\chi(u-\alpha)\,.
\end{align}
Using Eq.~\eqref{eq: wightman-holographic-reconstruction} with boundary smearing function $\varphi_\textsf{O},\varphi_\textsf{A}$, we get
\begin{align}
    \mathsf{W}_\skri(\varphi_\textsf{O},\varphi_\textsf{A})  
    &= -\frac{1}{4\pi^2}\lim_{\epsilon\to 0^+}\int\dd u\,\dd u'\,\frac{\chi(\frac{u}{T})\chi(\frac{u'-\alpha}{T})}{(u-u'-\ii\epsilon)^2}\notag \\
    &= \mathsf{W}(f_\textsf{O},f_\textsf{A})\,.
\end{align}
The second equality follows from the fact that the bulk unsmeared Wightman function in flat space reads
\begin{align}
    \mathsf{W}(\sx,\sy) &= -\frac{1}{4\pi^2}\frac{1}{(t-t'-\ii\epsilon)^2-|\bx-\by|^2}\,,
    \label{eq: Wightman-function-Minkowski}
\end{align}
thus the integral is (up to change of variable $u\to t$) exactly the bulk smeared Wightman function in Eq.~\eqref{eq: bulk-VEV-minkowski-gauss}.

\subsubsection{Case 2: spacelike pair $\emph{\textsf{OB}}$}

\begin{figure}[tp]
    \centering
    \includegraphics[scale=0.9]{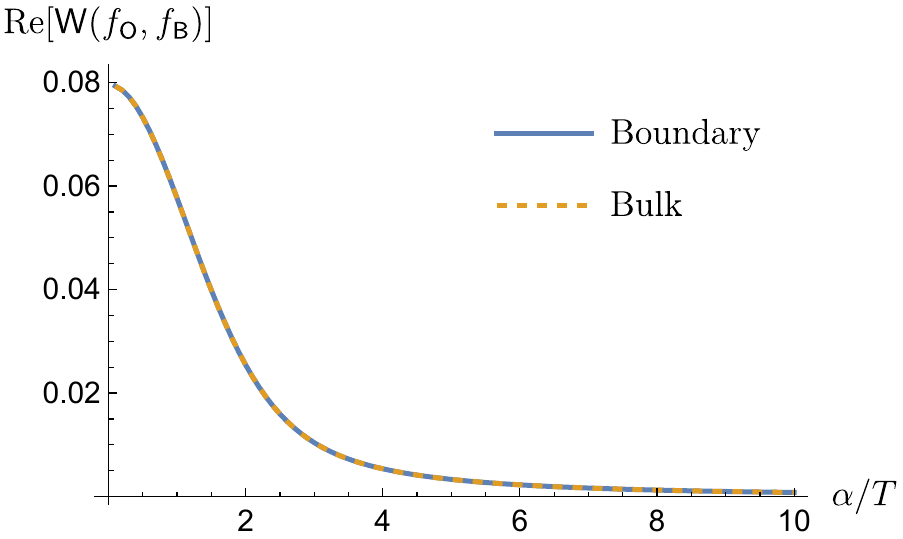}
    \caption{Real part of $\mathsf{W}(f_O,f_B)$ as a function of $\alpha/T$, where $\alpha = |\bx_\textsf{O}-\bx_{\textsf{B}}|$. The imaginary part vanishes since $\mathsf{OB}$ are spacelike separated.}
    \label{fig: spacelike-holography}
\end{figure}

For point $\textsf{B}$, we have $t_\textsf{B}=0$ and $\bx_\textsf{B}=(\alpha,0,0)$, and near $\skri^+$ the modified null variables are
\begin{align}
    u_j &= u + \alpha\cos\theta_\textsf{B}\,,\quad v_j = v - \alpha\cos\theta_\textsf{B}\,.
\end{align}
The boundary data is
\begin{align}
    \varphi_\textsf{B} &= -\frac{1}{4\pi }\chi\rr{\frac{u+\alpha\cos\theta_\textsf{B}}{T}}\,.
\end{align}
The boundary Wightman function therefore reads
\begin{align}
    \mathsf{W}_\skri(\varphi_\textsf{O},\varphi_\textsf{B}) 
    &= -\frac{1}{8\pi^2}\lim_{\epsilon\to 0^+}\int_0^\pi\dd\theta_\textsf{B}\sin\theta_\textsf{B}\notag\\
    &\hspace{0.5cm} \times \int\dd u\,\dd u'\,\frac{\chi(\frac{u}{T})\chi(\frac{u'+\alpha\cos\theta_\textsf{B}}{T})}{(u-u'-\ii\epsilon)^2}\,.
\end{align}
Using a change of variable $\tilde{u}' = u'+\alpha\cos\theta_\textsc{B}$ and integrating over $\theta_\textsc{B}$ first, we can rewrite this into a suggestive form
\begin{align}
    \mathsf{W}_\skri(\varphi_\textsf{O},\varphi_\textsf{B}) 
    &= -\frac{1}{4\pi^2}\lim_{\epsilon\to 0^+} \int\frac{\dd u\,\dd u'\,\chi(\frac{u}{T})\chi(\frac{u'}{T})}{(u-u'-\ii\epsilon)^2-|\alpha|^2}\,.
    \label{eq: spacelike-pair-points}
\end{align}
Let us first check this numerically (since in the FRW case we have to do this), as shown in Figure~\ref{fig: spacelike-holography}. Note that since the spacetime smearing is real and $\textsf{O}$ is spacelike separated from $\textsf{B}$, hence $\Im[\mathsf{W}(f_\textsc{O},f_\textsf{B})]=0$. Thus for spacelike pair of points we do get
\begin{align}
    \mathsf{W}_\skri(\varphi_\textsf{O},\varphi_\textsf{B}) = \mathsf{W}(f_\textsf{O},f_\textsf{B})\,.
\end{align}
We could obtain the exact expression using the fact that Eq.~\eqref{eq: spacelike-pair-points} has exactly the same esxpression as the smeared bulk Wightman function $\mathsf{W}(f_\mathsf{O},{f}_\mathsf{B})$ if we replace $t\to u$ and $|\bx-\by|^2=|\alpha|^2$ in Eq.~\eqref{eq: Wightman-function-Minkowski}.

Let us also remark that the form in Eq.~\eqref{eq: spacelike-pair-points} is highly suggestive, since  the same expression in Eq.~\eqref{eq: spacelike-pair-points} can be obtained by considering the final joint state of two Unruh-DeWitt qubit detectors interacting with massless scalar field at proper separation $\alpha$ for small/zero detector energy gap (the ``$\mathcal{L}_{\textsc{ab}}$'' term in the joint detector density matrix; see, e.g., \cite{pozas2015harvesting,Tjoa2021harvestingcomm}). Therefore these boundary correlators are in principle measurable by asymptotic observers who carry quantum-mechanical detectors. This is to be contrasted with the calculations done in, for instance, \cite{Wall2018asymptotic}, since it is not obvious how the correlators of the Bondi news tensor and Bondi mass can be measured in practice.

\subsubsection{Bulk reconstruction using \textit{smeared} Wightman functions}

It remains to show how to reconstruct the metric in the bulk. We will content ourselves with reconstructing $g_{tt}=-1$ and $g_{jj}=1$ at the origin $\sx=\mathsf{O}$ since we have translational invariance.

Due to modest holography, we have just seen that the bulk correlator $\mathsf{W}(f,g)$ \eqref{eq: bulk-VEV-minkowski-gauss} is also the expression for \textit{boundary correlator} $\mathsf{W}_\skri(\psi_f,\psi_g)$. Therefore our task is to simply reconstruct the metric using \eqref{eq: metric-reconstruction-boundary}. Through this prescription, the approximate expression for the metric component (denoted $g^{\delta,T}_{\mu\nu}$) at finite $T$ and $\delta$ are given by
\begin{widetext}
\begin{align}
    g^{\delta,T}_{tt}(\mathsf{O}) &\coloneqq \frac{T^2 \left(2 \sqrt{2} T e^{\frac{\delta ^2}{T^2}} \mathcal{F}\left(\frac{\delta }{\sqrt{2} T}\right)-\pi  \delta  \left[1+\text{erfi}\left(\frac{\delta }{\sqrt{2} T}\right)^2\right]\right)}{2 \delta  T e^{\frac{\delta ^2}{T^2}} \left[T-2 \sqrt{2} \delta  \mathcal{F} \left(\frac{\delta }{\sqrt{2} T}\right)\right]+\pi  \delta ^3 \left[1+\text{erfi}\left(\frac{\delta }{\sqrt{2} T}\right)^2\right]}\,, \quad 
    g^{\delta,T}_{jj}(\mathsf{O}) \coloneqq   \frac{T}{\sqrt{2} \delta  \mathcal{F}\left(\frac{\delta }{\sqrt{2} T}\right)}-\frac{T^2}{\delta ^2}\,.
    \label{eq: metric-reconstruction-smeared}
\end{align}
\end{widetext}
where {$\mathcal{F}(z)$ is the Dawson function and $\text{erfi}(z) = -\ii\,\text{erf}(\ii z)$ is defined from $\erf(z)$, the error function}. Now, keeping $\delta>0$ fixed but small for finite difference scheme, we have in the limit $T\to 0^+$ 
\begin{align}
    \lim_{T\to 0^+} g_{tt}^{\delta,T}(\mathsf{O}) = -1 \,,\quad \lim_{T\to 0^+} g_{jj}^{\delta,T}(\mathsf{O}) = 1\,.
\end{align}
hence we recover the non-trivial component of the Minkowski metric. It is important to note that the limits do not commute: we cannot, for instance, rescale $\Delta\coloneqq \delta/T$ and take $\Delta\to 0$. We need to keep $\delta$ finite or at least going to zero slower than $T$.

Note that if we start from the unsmeared \textit{bulk} Wightman function, we can \textit{easily} reconstruct the metric according to \cite{Kempf2016curvature,Kempf2021replace} because of the argument at the beginning of Section~\ref{sec: holographic-reconstruction} using the Hadamard form of the unsmeared Wightman function \eqref{eq: Hadamard-form}. For example, using the Wightman function \eqref{eq: Wightman-function-Minkowski}, it is straightforward to see that 
\begin{align}
    \mathsf{W}_\mathcal{M}(\sx,\sx')^{-1} &= -4\pi^2((t-t' )^2-(\bx-\bx')^2)
\end{align}
and hence by taking derivatives with respect to $\sx$ and $\sx'$ and dividing both sides by $-8\pi^2$ we simply get $g_{tt}=-1$ and $g_{jj}=1$ and $g_{\mu\nu}=0$ when $\mu\neq \nu$. However, for the boundary correlator, we cannot quite do this because there is no ``unsmeared'' version that is in the Hadamard form. We saw earlier in the calculation leading to Eq.~\eqref{eq: spacelike-pair-points} that for the spacelike pairs the boundary correlator $\mathsf{W}_\skri(\varphi_\mathsf{O},\varphi_\mathsf{B})$ may involve additional angular integral \textit{inside} the boundary smearing functions after propagating the bulk smearing functions $f_\mathsf{O},f_\mathsf{B}$ to $\skri^+$. This reflects the universal nature of $\skri^+$.

To summarize, our modest holographic reconstruction relies on two steps: (1) the bulk-to-boundary correspondence between the bulk and boundary correlators; (2) reconstructing the metric using \textit{smeared} boundary correlator. For Minkowski space, Step (2) can be done exactly, which is given in Eq.~\eqref{eq: metric-reconstruction-smeared}. In the next example for FRW spacetimes, Step 2 will be numerically difficult to compute, so we will content ourselves with making sure Step 1 is achieved and Step 2 follows in analogous fashion as Minkowski space using prescription \eqref{eq: metric-reconstruction-boundary}.

\subsection{Example 2: FRW spacetime}

\begin{figure*}[tp]
    \centering
    \includegraphics[scale=0.8]{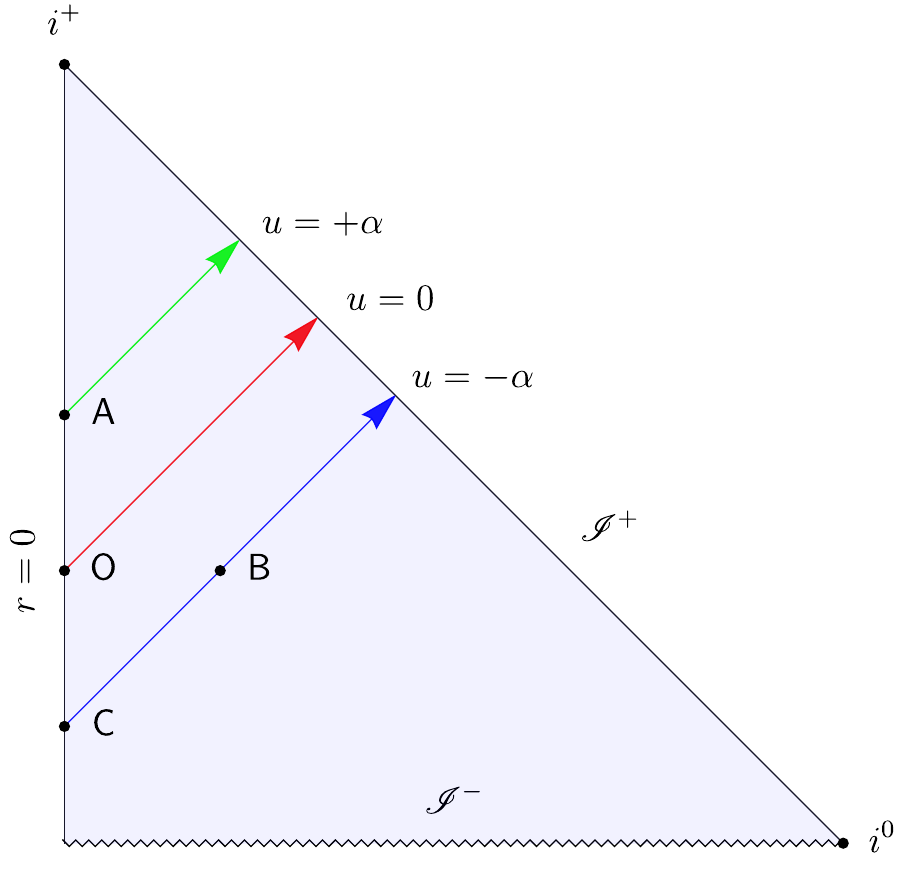}
    \includegraphics[scale=0.8]{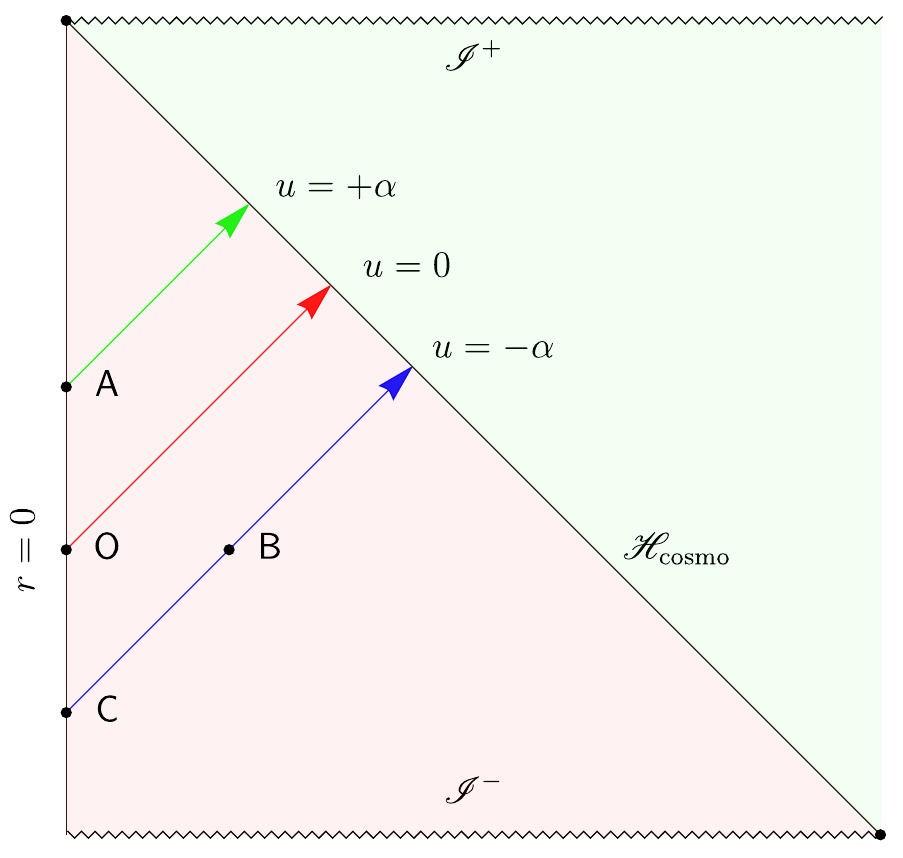}
    \caption{Penrose diagram for the bulk-to-boundary reconstruction in FRW spacetime. \textit{Left:} spatially flat FRW geometry with zero cosmological constant and dust/radiation matter content. \textit{Right:} spatially flat FRW geometry with positive cosmological constant describing Big Crunch (left red patch) or Big Bang (right green patch). Both the past/future conformal infinity or initial/final singularity $\mathscr{I}^\pm$ are spacelike and $\mathscr{H}_{\text{cosmo}}$ is the past/future cosmological horizon.}
    \label{fig: penrose-reconstruction-FRW}
\end{figure*}

The FRW universe with flat spatial section is given by the line element
\begin{align}
    \dd s^2
    &= -\dd t^2+a(t)^2(\dd r^2+r^2\dd\Omega^2)
\end{align}
where $a(t)$ is the scale factor and the spatial section is written in spherical coordinates. It is convenient to recast this metric into the conformally flat form by using conformal time $\eta = \int^t\dd t'/a(t')$, so that the metric reads
\begin{align}
    \dd s^2 = a(\eta)^2(-\dd \eta^2+\dd x^2+\dd y^2+\dd z^2)\,.
\end{align}
Here we have used the Cartesian coordinates for the spatial section which is convenient when computing the Wightman function. We will use the spherical coordinates when we calculate the projection to the null boundary.

The bulk Wightman function is conformally related to Minkowski one by the relation~\cite{birrell1984quantum}
\begin{align}
    \mathsf{W}_{\textsc{FRW}}(\sx,\sy) = a^{-1}(\eta_\sx)\mathsf{W}_{\textsc{M}}(\sx,\sy)a^{-1}(\eta_\sy)\,,
    \label{eq: Weyl-rescaling-correlator}
\end{align}
where $a(\eta_\sx)$ is the scale factor evaluated at point $\sx$. It follows that the unsmeared Wightman function reads
\begin{align}
    \mathsf{W}_{\textsc{FRW}}(\sx,\sy) = -\frac{1}{4\pi^2}\frac{a(\eta_\sx)^{-1}a(\eta_\sy)^{-1}}{(\Delta\eta-\ii\epsilon)^2-|\Delta\bx|^2}\,,
    \label{eq: Wightman-function-FRW}
\end{align}
where $\Delta\eta=\eta_\sx-\eta_\sy$ and $\Delta\bx =\bx-\by$.
In what follows we will drop the subscript $\textsc{FRW}$ to remove clutter.

If we regard the spacetime smearing as being associated to observers prescribing the interaction in \textit{comoving time} $t$, then the we can consider similar pointlike function
\begin{align}
    f_j(\sx) &=\chi\rr{\frac{t(\eta)-t_j}{T}}\delta^3(\bx-\bx_j)
    \,,
\end{align}
where now $t(\eta)$ is written as a function of conformal time. The bulk smeared Wightman function is thus given by 
\begin{align}
    &\mathsf{W}(f_i,f_j) \notag\\
    &= -\frac{1}{4\pi^2}\lim_{\epsilon\to 0}\int \frac{\dd\eta\,\dd\eta'}{a(\eta)^{-3}a(\eta')^{-3}}\frac{\chi(t(\eta)-t_i)\chi(t(\eta')-t_j)}{(\eta-\eta'-\ii\epsilon)^2-|\Delta\bx_{ij}|^2}\,.
\end{align}
As before, we need the causal propagator to find the boundary correlator. The causal propagator is obtained using the Weyl rescaling in Eq.~\eqref{eq: Weyl-rescaling-correlator}, so that it reads
\begin{align}
    E(\sx,\sx') &= \frac{\delta(\Delta\eta+|\Delta\bx|)-\delta(\Delta\eta-|\Delta\bx|)}{4\pi\, a(\eta)a(\eta')|\Delta\bx|}\,.
\end{align}
We can then define a set of modified null coordinates
\begin{equation}
    \begin{aligned}
    U_j &= \eta - |\bx-\bx_j|\,,\\
    V_j &= \eta + |\bx-\bx_j|\,.
\end{aligned}
\label{eq: modified-null-coord-FRW}
\end{equation}
It follows that
\begin{align}
    Ef_j(\sx) &= \int \dd^4\sy\sqrt{-g} \,E(\sx,\sy)f_j(\sy)\notag\\
    &= \frac{a(V_j)^3\chi\rr{\frac{t(V_j)-t_j}{T}}- a(U_j)^3\chi\rr{\frac{t(U_j)-t_j}{T}}}{4\pi\,a(\eta) |\bx-\bx_j|}\,.
    \label{eq: Efj-bulk-FRW}
\end{align}
The boundary data associated to $Ef_j$, denoted by $\varphi_j$, is is the projection of $Ef_j$ to $\skri^+$ via the projection map $\Gamma$. In this limit, the modified null variables become
\begin{equation}
    \begin{aligned}
    U_j &\to u_j\coloneqq u + |\bx_j|\cos\theta_j \,,\\
    V_j &\to v_j\coloneqq v - |\bx_j|\cos\theta_j \,,
    \label{eq: modified-null-coord-boundary-FRW}
    \end{aligned}
\end{equation}
where $\theta_j$ is the angle between $\bx$ and $\bx_j$. More concretely, the projection map amounts to rescaling by $\Omega^{-1} = r a(\eta)$, taking $r=|\bx|\to\infty$ and keeping $u=\eta-r$ fixed, i.e.,
\begin{align}
    \Gamma Ef = \lim_{\skri^+} r a(\eta) Ef\,.
\end{align}
From this we get
\begin{align}
    \varphi_j(u,x^A) &= -\frac{a(u_j)^3}{4\pi}\chi\rr{\frac{t(u_j)-t_j}{T}}\,.
    \label{eq: boundary-data-FRW}
\end{align}

In order to make explicit calculations, we need to use a concrete scale factor. For our purposes, we are interested in physically relevant scale factor $a(t)$ associated to perfect fluid stress-energy tensor
\begin{align}
    T_{\mu\nu} &= (\rho+p)\mathsf{u}_\mu \mathsf{u}_\nu +p g_{\mu\nu}\,,
\end{align}
where $\mathsf{u}^\mu$ is the four-velocity of the fluid, $\rho$ and $p$ are the energy density and pressure (as a function of only the comoving/conformal time). The fluid is assumed to obey the barotropic equation of state $p = (\gamma-1)\rho$, where $0\leq \gamma \leq 2$. The conservation law $\nabla_\mu T^{\mu\nu}=0$ implies that the evolution of $\rho,p$ are constrained to obey
\begin{align}
    \frac{\dot\rho}{\rho+p} = -3\frac{\dot{a}}{a}\,.
\end{align}
This implies, in particular, that
\begin{align}
    \rho = \frac{\rho_0}{a^{3\gamma}}
\end{align}
where $\rho_0$ is some constant. For dust-filled universe, we have $\gamma=1$ so $\rho\propto a^{-3}$ and for radiation-filled universe we have $\gamma = 4/3$ so $\rho\propto a^{-4}$. The value $\gamma=0$ corresponds to de Sitter universe with cosmological constant $\Lambda>0$ by setting $\Lambda = \rho_0$ (see \cite{griffiths2009exact} for more details on FRW geometry). 

\begin{figure*}
    \centering
    \includegraphics[scale=0.9]{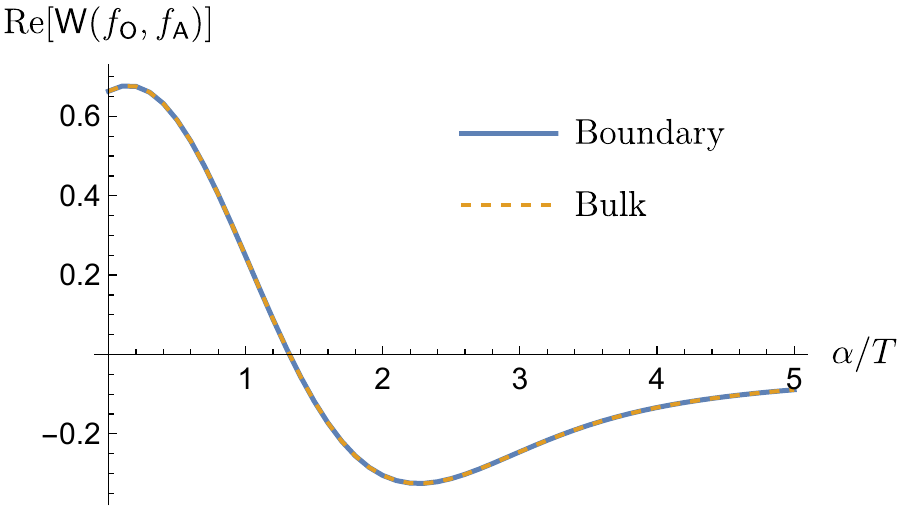}
    \includegraphics[scale=0.9]{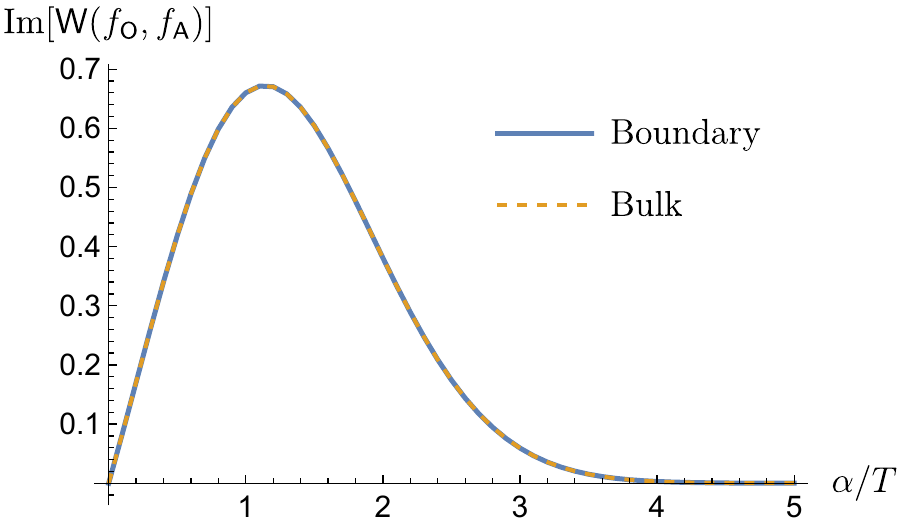}
    \includegraphics[scale=0.9]{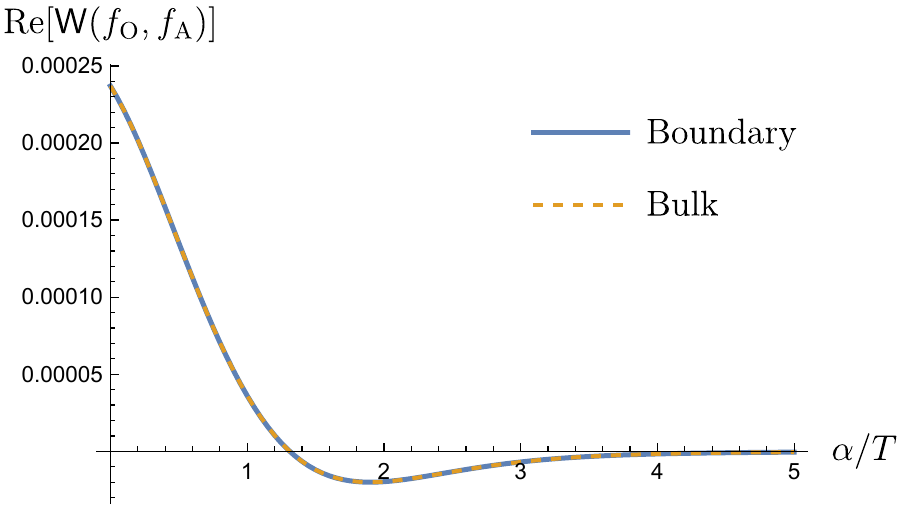}
    \includegraphics[scale=0.9]{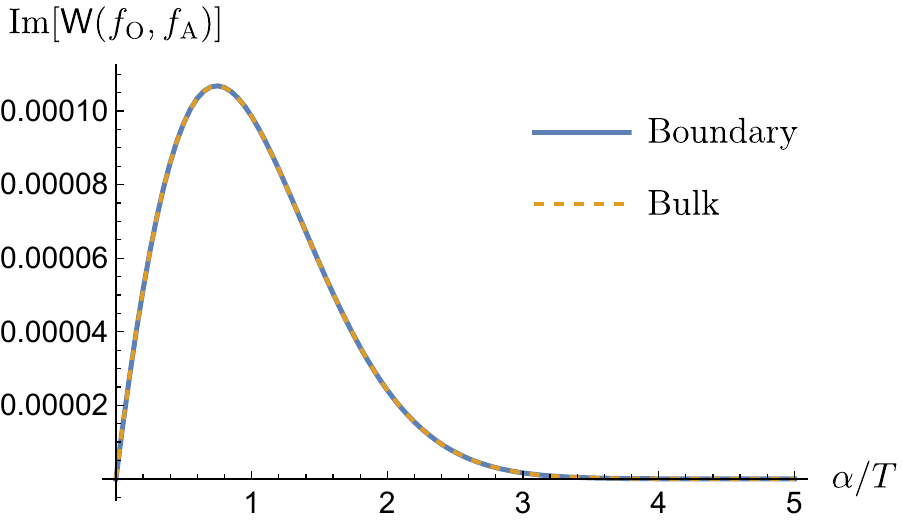}
    \caption{Real and imaginary parts of $\mathsf{W}(f_\textsf{O},f_\textsf{A})=\mathsf{W}_\skri(\varphi_\mathsf{O},\varphi_\mathsf{A})$ in FRW spacetime. We pick $HT=0.2$ and we plot against $\alpha = (t_\textsf{O}-t_\textsf{A})/T$. \textit{Top row:} radiation-dominated universe with $a(\eta)=H\eta$ and $\eta>0$. \textit{Bottom row:} de Sitter {contracting} universe with $a(\eta) = (H\eta)^{-1}$ and $\eta>0$.}
    \label{fig: timelike-holography-FRW}
\end{figure*}

Under the above assumptions of the matter content in the bulk geometry, the corresponding scale factors for these two classes of FRW spacetimes are given by 
\begin{align}
    a_{\gamma}(\eta) = (H\eta)^{\frac{2}{3\gamma-2}}\,,\quad a_{\text{dS}}(\eta) &= \pm \frac{1}{H\eta}\,,
\end{align}
where $H>0$ is a constant with dimension of inverse time. The Penrose diagram for the respective classes of FRW geometries are shown in Figure~\ref{fig: penrose-reconstruction-FRW}. For concreteness, we will restrict our attention to (a) radiation-filled universe $\gamma=4/3$ with $a(\eta) = H\eta$  and $\eta>0$; (b) \textit{contracting} de Sitter universe with $a(\eta) = 1/(H\eta)$ and $\eta>0$. The reason we include the de Sitter universe is to highlight one non-trivial aspect of this construction: that is, even if the spacetime is asymptotically de Sitter {and $\skri^+$ is spacelike}, the \textit{cosmological horizon} $\mathscr{H}_\text{cosmo}$ {shares analogous features}
\footnote{It is important to note that the symmetry group of the horizon is distinct from the BMS group but a careful treatment~\cite{Dappiaggi2008cosmological} shows that the cosmological horizon's algebraic state $\omega_{\mathscr{H}_\text{cosmo}}$ is invariant under exactly these transformations.} as future null infinity $\skri^+$ for asymptotically flat spacetimes  \cite{Dappiaggi2008cosmological}.

As before, due to spherical symmetry we only need to attempt the reconstruction of the bulk Wightman function for three types of pair of events, which we label by the same points $\textsf{O,A,B,C}$. In the conformal coordinates $(\eta,r,x^A)$, setting $x^A=0$ by spherical symmetry.

\subsubsection{Timelike pairs $\emph{\textsf{OA}}$}

Let us take $\textsc{O} = (\eta(t_\textsc{O}),0,0,0)$, $\textsc{A} = (\eta(t_\textsc{A}),0,0,0)$, where $t_\textsc{O}$ and $t_\textsc{A}$ are some positive constants.  From Eq.~\eqref{eq: boundary-data-FRW}, we have
\begin{align}
    \varphi_\textsc{O}(u,x^A) 
    &= -\frac{a(u)^3}{4\pi}\chi\rr{\frac{t(u)-t_\textsc{O}}{T}}\,,\\
    \varphi_\textsc{A}(u,x^A) 
    &= -\frac{a(u)^3}{4\pi}\chi\rr{\frac{t(u)-t_\textsc{A}}{T}}\,.
\end{align}
and we define $\alpha = t_\textsf{A}-t_\textsf{O}$. For radiation and de Sitter scale factors, the comoving time $t$ is given in terms of conformal time by
\begin{subequations}
\begin{align}
    t_{\text{rad}}(\eta) &= \frac{H\eta^2}{2}\,,\quad \eta>0\\
    \quad t_\text{dS}(\eta) &= H^{-1}\log(H\eta) \,,\quad \eta > 0\,.
\end{align}
\end{subequations}

Now we can compute the boundary correlator
\begin{align}
    \mathsf{W}_\skri(\varphi_\textsf{O},\varphi_\textsf{A})
    &= -\frac{1}{4\pi^2}\lim_{\epsilon\to 0^+}\int\dd u\,\dd u'\,a(u)^3a(u')^3\notag\\
    &\hspace{1.5cm}\times \frac{\chi(\tfrac{t(u)}{T})\chi(\tfrac{t(u')-\alpha}{T})}{(u-u'-\ii\epsilon)^2}\,.
\end{align}
The results are shown in Figure~\ref{fig: timelike-holography-FRW} for both the radiation-dominated universe and the de Sitter contracting universe. We see that they clearly agree. However, observe that $\chi(t(u)/T)$ is \textit{not} Gaussian and the supports of $\varphi_j$ can be quite different. For example, in the de Sitter contracting universe case $\varphi_j$ is a smooth function with support only on the positive real axis, i.e., $\supp (\psi_j)\subset (0,\infty)$. The takeaway is that different bulk geometries are accounted for by different ``boundary data'' at the conformal boundary, in this case either $\skri^+$ or $\mathscr{H}_\text{cosmo}$.

\subsubsection{Spacelike pairs $\emph{\textsc{OB}}$}

\begin{figure*}[tp]
    \centering
    \includegraphics[scale=0.9]{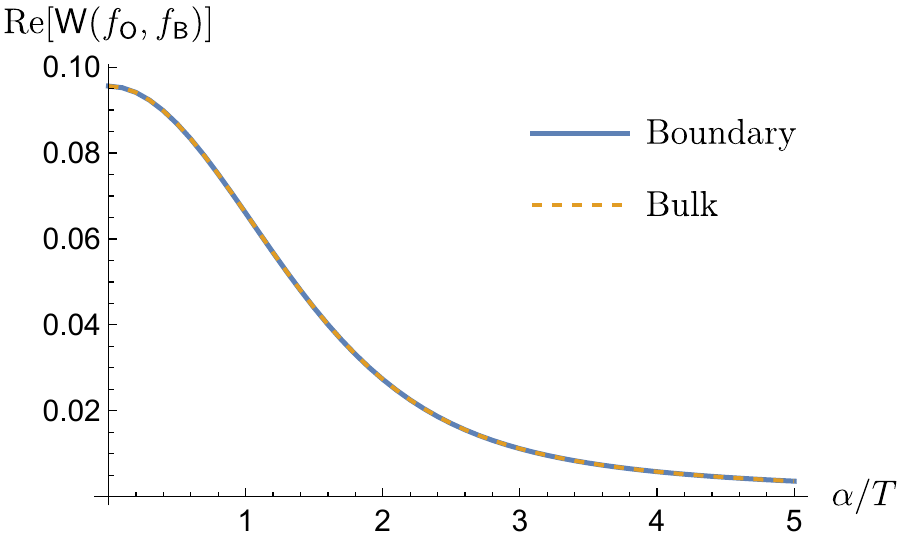}
    \includegraphics[scale=0.9]{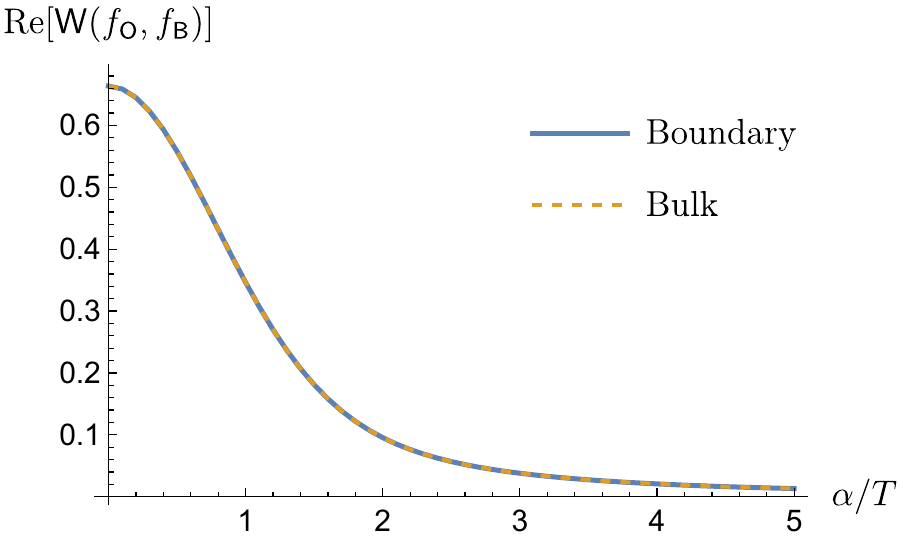}
    \caption{Real part of $\mathsf{W}(f_\textsf{O},f_\textsf{B})=\mathsf{W}_\skri(\varphi_\textsf{O},\varphi_\textsf{B})$ in FRW spacetimes (imaginary part vanishes). We pick $HT=0.2$ and we plot against $\alpha = 
    |\bx_\textsf{O}-\bx_{\textsf{B}}|/T$. \textit{Left}: radiation-dominated universe. \textit{Right}: de Sitter contracting universe.}
    \label{fig: spacelike-holography-FRW}
\end{figure*}

Let us take $\textsf{O} = (\eta(t_\textsf{O}),0,0,0)$, $\textsf{B} = (\eta(t_\textsf{O}),\alpha,0,0)$, where $t_\textsf{O}$ is some fixed constant chosen so that $\textsf{O},\textsf{B}$ are on the same time slice and $t_\textsc{A} = \alpha>0$. From Eq.~\eqref{eq: boundary-data-FRW}, we have
\begin{align}
    \varphi_\textsc{B}(u,x^A) 
    &= -\frac{a(u+\alpha \cos\theta)^{3}}{4\pi}\chi\rr{\frac{t(u)}{T}}\,.
\end{align}
This time we have an angular integral, so the boundary correlator reads
\begin{align}
    &\mathsf{W}_\skri(\varphi_\textsf{O},\varphi_\textsf{B}) \notag\\
    &= -\frac{1}{8\pi^2}\lim_{\epsilon\to 0^+}\int \dd u\,\dd u'\int \sin\theta\dd\theta\,a(u)^3a(u'+\alpha\cos\theta)^3\notag\\
    &\hspace{2cm}\times\frac{\chi(\frac{t(u)}{T})\chi(\frac{t(u'+\alpha\cos\theta)}{T})}{(u-u'-\ii\epsilon)^2}\,.
\end{align}
By change of variable $\tilde{u}' = u' + \alpha\cos\theta$ and integrating over $\theta$, the boundary correlator can be simplified into 
\begin{align}
    &\mathsf{W}_\skri(\varphi_\textsf{O},\varphi_\textsf{B}) \notag\\
    &= -\frac{1}{4\pi^2}\lim_{\epsilon\to 0^+}\int \dd u\,\dd u'\,
    \frac{a(u)^3a(u')^3 \chi(\frac{t(u)}{T})\chi(\frac{t(u')}{T})}{(u-u'-\ii\epsilon)^2-|\alpha|^2}\notag\\
    &= \mathsf{W}(f_\textsf{O},f_\textsf{B})\,.
    \label{eq: spacelike-correlator-boundary}
\end{align}
The second equality is obtained simply by comparing with the bulk Wightman function expression, since $\Delta\bx_{\textsc{OB}}=\alpha$. The results are shown in Figure~\ref{fig: spacelike-holography-FRW}.

\subsubsection{Holographic reconstruction of the bulk FRW spacetimes}

We have shown that the bulk-to-boundary correspondence of the correlators work as well in FRW spacetimes. The scheme works for the asymptotically flat radiation-dominated universe where the ``conformal boundary'' is future null infinity $\skri^+$ as expected. A nice bonus is that, as shown in \cite{Dappiaggi2008cosmological}, the same construction ought to work as well for the de Sitter contracting universe. However, in this case the bulk-to-boundary correspondence is not between the bulk and the conformal boundary $\skri^+$ (which is \textit{spacelike}), but rather with the cosmological horizon $\mathscr{H}_\text{cosmo}$ (\textit{c.f.} Figure~\ref{fig: penrose-reconstruction-FRW}). Hence for de Sitter cosmological spacetime it is perhaps a misnomer to call it bulk-to-boundary correspondence. However, since the cosmological horizon is also a codimension-1 null hypersurface, we still have holographic reconstruction of bulk geometry from ``boundary data''. 

In principle, the metric can be reconstructed analogous to the procedure outlined for Minkowski space using Eq.~\eqref{eq: metric-reconstruction-boundary}, but because the boundary Wightman function does not admit simple closed-form expression, it is difficult to perform this calculation numerically since we need $T,\delta$ to be very small. However, it is worth noting that there is something universal about the boundary correlator: take, for instance, the case when the two points are spacelike in Eq.~\eqref{eq: spacelike-correlator-boundary} which we reproduce for convenience:
\begin{align}
    &\mathsf{W}_\skri(\varphi_\textsf{O},\varphi_\textsf{B}) \notag\\
    &= -\frac{1}{4\pi^2}\lim_{\epsilon\to 0^+}\int \dd u\,\dd u'\,
    \frac{a(u)^3a(u')^3 \chi(\frac{t(u)}{T})\chi(\frac{t(u')}{T})}{(u-u'-\ii\epsilon)^2-|\alpha|^2}\,.\notag
\end{align}
This integral differs from the one in Minkowski space (\textit{c.f.} Eq.~\eqref{eq: spacelike-pair-points}) only in the choice of boundary smearing functions and the physical meaning of $|\alpha|$: in Minkowski space, it amounts to setting $a(u) = 1$ and $t(u) = u$. Therefore, information of the bulk geometry is encoded in the  boundary data (smearing) that enters into this ``universal integral'' over $u,u'$ and angular variable $x^A$. 

{The fact that the boundary smearing functions contain information about the geometry cannot be understated. In particular, one cannot  ``cheat'' by trying to reconstruct the bulk metric from \textit{unsmeared bulk correlator}.} If we use the unsmeared bulk correlator \eqref{eq: Wightman-function-FRW}, we can check that
\begin{align}
    \hspace{-0.2cm}\lim_{\sx'\to\sx}\partial_\mu\partial_{\nu'}\mathsf{W}(\sx,\sx')^{-1}
    &= 
    \begin{cases}
    +{8\pi^2}{a(\eta)^2}\quad \mu=\nu'= 0 \\
    -{8\pi^2}{a(\eta)^2}\quad \mu=\nu'= j \\
    0\quad\quad\quad \quad \text{otherwise}
    \end{cases}
\end{align}
so that indeed the metric components are $g_{\mu\nu}(\sx) = \mp a(\eta)$ for $\mu=\nu=0$ and $\mu=\nu=j$ respectively (and zero otherwise). This works because of the Hadamard form of the (unsmeared) Wightman function \eqref{eq: Hadamard-form}. We cannot quite do this literally for the boundary correlator because the ``unsmeared'' part is universal: as we will see in the next section, it has the structure of
\begin{align}
    \mathsf{W}_\skri(u,u',x^A,y^A) \sim -\frac{1}{\pi}\frac{\delta_{S^2}(x^A-y^A)}{(u-u'-\ii\epsilon)^2}\,, 
\end{align}
where $\delta_{S^2}(x^A-y^A)$ is the Dirac delta distribution on two-sphere. This universality is a manifestation of the universality of $\skri^+$ (or $\mathscr{H}_\text{cosmo}$ for de Sitter case).

\section{Asymptotic expansion of the field operator}
\label{sec: large-r-expansion}

We should mention that the projection $\Gamma$ acting on the space of solutions $\Sol_\R(\M)$ could also be viewed at the level of canonical quantization. This is what is typically done in the {``infrared triangle''} program \cite{strominger2018lectures,he2014new,pasterski2017asymptotic}, where the idea is to perform \textit{asymptotic large-$r$ expansion} of the field operator and keeping only the leading term. This way of thinking is highly intuitive because it does not require us to think of unphysical spacetime $\tilde{\M}$ and it compels us to think of scalar QFT at $\skri^+$ to be an approximation of ``faraway observers''. The price to pay is that the holographic nature of the QFT degrees of freedom is not obvious because $\skri^+$ is not strictly speaking part of the description by faraway observers {(since they travel on timelike curves)}. 

Let us now show how the two methods are related, using Minkowski space example as a reference, and connect the holographic nature of the QFT to asymptotic observers. This connection implies that QFT at $\skri^+$ \textit{can} and \textit{should} be accessible to physical asymptotic (large-$r$) observers.

First, for Minkowski spacetime the canonical quantization gives the ``unsmeared'' field operator
\begin{align}
    \hat\phi(\sx) &= \int\frac{\dd^3\bk}{\sqrt{2(2\pi)^3|\bk|}}\hat a_\bk e^{-\ii|\bk|t+\ii\bk\cdot\bx} + \text{h.c.}\,.
\end{align}
It is useful to write this in Bondi chart $\sx = (u,r,x^A)$. Using the fact that the metric in Bondi coordinates is given by
\begin{align}
    \dd s^2 &= -\dd u^2 -2\dd u\,\dd r+ r^2\dd\Omega^2\,,
\end{align}
we have $ k_\mu x^\mu = -\omega u - \omega r (1-\hat\bk\cdot \hat\br)$, where $\omega=|\bk|$, $\hat\bk = \bk/|\bk|$ and $\hat\br = \br/|\br|$ are unit vectors. We can then write $\hat\bk\cdot\hat\br = \cos\theta$ for some angle $\theta$ and {$\dd^3\bk=\omega^2\dd\omega \,\dd\gamma_{S^2} $}. The field operator now reads
\begin{align}
    \hat\phi(u,r,x^A) &= \frac{1 }{\sqrt{2(2\pi)^3}}\int_0^\infty \!\!\!\omega^\frac{3}{2}\dd\omega\notag\\
    &\times \int{\dd \gamma_{S^2}}\,\hat a_\bk e^{-\ii|\bk|u-\ii\omega r(1-\cos\theta )}+\text{h.c.}\,.
\end{align}
Now we would like to take large-$r$ limit. The stationary phase approximation says that for any function $f(\bk)$ we have
\begin{align}
    \int{\dd \gamma_{S^2}} f(\bk)e^{\pm \ii|\bk| r(1-\hat\bk\cdot\hat\br)} \sim \pm\frac{2\pi \ii}{|\bk|r}f(|\bk|\hat r) + \mathcal{O}(r)\,.
\end{align}
This implies that at leading order in $r$ the field operator is dominated by
\begin{align}
    \hat\phi(u,r,x^A) &\sim  -\frac{\ii}{2r\sqrt{\pi}}\int_0^\infty\!\!\!\!\!\omega^{\frac{1}{2}} \dd\omega  \Bigr[\hat a_{\omega\hat r}e^{-\ii\omega u} -\hat a^\dagger_{\omega\hat r}e^{\ii \omega u}\Bigr]\,.
\end{align}
The \textit{boundary data} (unsmeared) operator is then defined to be
\begin{align}
    \hat\varphi(u,x^A)\coloneqq \lim_{r\to\infty }r\hat\phi(u,r,x^A)\,,
    \label{eq: boundary-dat field}
\end{align}
and the creation operators satisfy the following commutation relation
\begin{equation}
    \left[\hat a^{\phantom{\dagger}}_{\omega\hat r},\hat a^\dagger_{\omega'\hat r'}\right]=\frac{\delta(\omega-\omega')}{\omega^2}\delta_{S^2}(\hat r-\hat r')\,.
\end{equation}

Let us now compute the (unsmeared) Wightman two-point function at $\skri^+$ with respect to the vacuum state\footnote{This vector state $\ket{0_\skri}$ is obtained from the $\mathsf{BMS}_4$-invariant algebraic state $\omega_\skri$ via GNS representation theorem.}. One important subtlety arises here---from dimensional analysis and scaling arguments it can be seen the ordinary Wightman function, $\braket{0_\skri|\hat\varphi(u,x^A)\hat\varphi(u',y^A)|0_\skri}$, is logarithmically divergent at $\skri^+$. Thus instead we compute the two-point correlators of its conjugate momentum $\partial_u\hat\varphi$:
\begin{align}
    &\mathsf{W}_\skri(u,x^A; u',y^A) = \braket{0_\skri|\partial_u\hat\varphi(u,x^A)\partial_u\hat\varphi(u',y^A)|0_\skri}\notag\\
    &= \int\!\!\frac{\dd\omega\dd\omega'}{4\pi}{\dd \gamma_{S^2}}\, {\dd \gamma'_{S^2}}(\omega\omega')^{3/2}e^{-\ii\omega u-\ii\omega'u'}\notag\\
    &\qquad\times\braket{0_\skri|\hat a_{\omega\hat r}^{\phantom{\dagger}}\hat a_{\omega'\hat r'}^\dagger|0_\skri} \notag\\
    &= \frac{1}{4\pi}\int\omega\,\dd\omega\,{\dd \gamma_{S^2}}\, {\dd \gamma'_{S^2}} e^{-\ii\omega (u-u')} \delta_{S^2}(\hat\br-\hat\br') \notag\\
    &= -\frac{1}{4\pi}\lim_{\epsilon\to 0}\frac{1}{(u-u'-\ii\epsilon)^2}\int{\dd \gamma_{S^2}}\,{\dd \gamma'_{S^2}} \delta_{S^2}(\hat\br-\hat\br')\,.
\end{align}
Now if we integrate this over smearing functions $\Psi_1,\Psi_2$ at $\skri^+$, we get
\begin{align}
    &\mathsf{W}_\skri(\Psi_1,\Psi_2) \notag\\
    &=- \frac{1}{4\pi}\lim_{\epsilon\to 0}\!\int\!\dd u\,\dd u'{\dd \gamma_{S^2}}\frac{\Psi_1(u,x^A)\Psi_2(u',x^A)}{(u-u'-\ii\epsilon)^2}\,,
    \label{eq: boundary-Wightman-asymptotic}
\end{align}
where we use capital Greek letter $\Psi$ to distinguish it with the boundary smearing function $\psi\in \Sol_\R(\skri^+)$ in AQFT approach. 

Observe that Eq.~\eqref{eq: boundary-Wightman-asymptotic} appears to be off by a factor of $1/4$ compared to Eq.~\eqref{eq: boundary-Wightman} obtained using algebraic method. This discrepancy arises because the algebraic approach calculates this two-point function somewhat differently. {To see this, note that for $\psi\in \Sol_\R(\skri^+)$ the smeared boundary field operator $\hat\varphi(\psi)$ is related to the unsmeared one via \textit{symplectic} smearing, i.e., we want to define $\hat\varphi(\psi)\,\,``\!\!\coloneqq\!\!"\,\,\sigma_\skri(\psi,\hat\varphi)$. However, by using integration by parts on Eq.~\eqref{eq: skri-symplc}, we get
\begin{align}
    \sigma_\skri(\psi,\hat\varphi)
    &= {2}\int_{\skri^+}\!\!\!\!\dd u\,{\dd \gamma_{S^2}}\,\psi(u,x^A)\partial_u\hat\varphi(u,x^A)\notag\\
    &\equiv 2\int_{\skri^+}\!\!\!\!\dd u\,{\dd \gamma_{S^2}}\,\psi(u,x^A)\hat\Pi(u,x^A)\notag\\
    &\eqqcolon 2\,\hat{\Pi}(\psi)\,.
\end{align}
where $\hat\Pi(u,x^A)=\partial_u\hat\varphi(u,x^A)$ is the (unsmeared) conjugate momentum to $\hat\varphi(u,x^A)$. Hence the unsmeared boundary field operator $\hat\varphi(\psi)$ should be interpreted as the \textit{smeared} conjugate momentum operator $\partial_u\hat\varphi$, not the smeared boundary field operator $\hat\varphi$ itself. Note that in null surface quantization, the operator $\Pi$ is \textit{not} independent of $\hat\varphi$ \cite{frolov1978null}, unlike in the bulk scalar theory.}

The holographic reconstruction works by fixing $f\in \CS$, propagate it to $\skri^+$ by taking
\begin{align}
    \psi_f(u,x^A) \coloneqq \lim_{r\to\infty} (\Omega^{-1} Ef)(u,r,x^A)
\end{align}
and calculating
\begin{align}
    \mathsf{W}_\skri(\psi_f,\psi_g) &= -\frac{1}{\pi}\lim_{\epsilon\to 0}\int {\dd \gamma_{S^2}}\dd u\,\dd u'\frac{\psi_f(u,x^A)\psi_g(u',x^A)}{(u-u'-\ii\epsilon)^2}\,.
\end{align}
Since Eq.~\eqref{eq: boundary-Wightman-asymptotic} is based on interpretation of {smeared conjugate momentum operator}
\begin{align}
    \hat\Pi(\Psi) = \int_{\skri^+}\dd u\,{\dd \gamma_{S^2}}\,\Psi(u,x^A){\partial_u}\hat\varphi(u,x^A)\,,
\end{align}
{this means that $\psi_f$ that appear directly in Eq.~\eqref{eq: boundary-Wightman} is related to \textit{symplectic smearing} $\psi$ in $\hat\varphi(\psi)$ and ``momentum smearing'' $\Psi$ in $\hat\Pi(\Psi)$} by
\begin{align}
    \psi_f = \psi = \frac{\Psi}{2} \in \Sol_\R(\skri^+)\,.
\end{align}
The key takeaway is that the smeared Wightman two-point functions computed using algebraic approach and large-$r$ expansion of the bulk (unsmeared) field operator only differ by a normalisation.

\section{Discussion and outlook}
\label{sec: discussion}

In this work, we have shown that one can directly reconstruct the bulk geometry of asymptotically flat spacetimes from the boundary correlators at $\skri^+$. This makes use of two previously unconnected results: augmenting the bulk-to-boundary correspondence developed in the AQFT community~\cite{Dappiaggi2005rigorous-holo,Dappiaggi2008cosmological,Dappiaggi2009Unruhstate,dappiaggi2015hadamard} with the recent metric reconstruction method using scalar correlators based on \cite{Kempf2016curvature,Kempf2021replace}. The version that is more relevant for us is the scheme used in \cite{perche2021geometry} is more appropriate due to the more direct use of Wightman two-point functions. This makes explicit use of the uniqueness and Hadamard nature of the boundary field state and importantly is relevant for asymptotic observers. The idea is that while no physical observers can follow null geodesics exactly on $\skri^+$, we can perform a large-$r$ expansion of bulk field operator. The asymptotic observers near $\skri^+$ will thus find that the bulk correlation functions $\mathsf{W}_\M(f,g)$ very close to $\skri^+$ is at leading order given exactly by $\mathsf{W}_\skri(\psi_f,\psi_g)$. 

We perform our calculations for relatively simple examples, namely both Minkowski and FRW spacetimes, where we can show concretely how the boundary smeared correlators have universal structure (reflecting the universal structure of $\skri^+$) and much of the geometric information is encoded in the boundary smearing functions, i.e. boundary data. Furthermore, the calculations are explicit enough for us to see that the boundary correlators can be expressed in the language of Unruh-DeWitt (UDW) detectors used in relativistic quantum information (RQI).

That is, for asymptotic observers who carry qubit UDW detectors interacting with a massless scalar field, the expressions for the boundary correlators naturally appear in the final density matrix of the detectors (see, e.g., \cite{Tjoa2021harvestingcomm,pozas2015harvesting}). In terms of detectors, the differences between Minkowski and FRW scenarios manifest as different ``switching functions'' (i.e., different interaction profiles). Therefore, the holographic reconstruction can be properly expressed in operational language using tools from RQI, since the correlators can indeed be extracted directly via quantum state tomography, without assuming that any correlators are simply ``measurable''.

There are several future directions now to explore within this framework. First, concretely understanding the projection map $\Gamma$ in generic spacetimes seems difficult, {since one needs to have a very good handle on causal propagators $E(\sx,\sx')$.} However, by making use of Bondi coordinates (e.g. Eq.~\eqref{eq: Bondi metric}) one may be able to systematically construct the asymptotic expansion of the causal propagator and see the radiative data of the gravitational field directly in the boundary correlators for asymptotic observers. 

{For example, one may wonder if boundary correlators may have imprints that can be used to infer the existence of gravitational (shock)waves \cite{Smith2020harvestingGW,Dray1985shockwave}, since the bulk correlators know about the background shockwave (see, e.g., \cite{Gray2021imprint})}. {On the other hand}, recently complex calculations of bulk correlators have become possible for Schwarzschild spacetimes and even the interior of Kerr spacetime (see, e.g., \cite{Casals2020commBH,zilberman2022two}). Modest holography suggests that near-horizon and near-$\skri$ correlators \cite{Dappiaggi2009Unruhstate} can perhaps aid in these fronts, in which case one can then reconstruct the black hole geometry from near-horizon and asymptotic correlators.

{Second, a natural extension of this construction is to see whether the result generalizes to massive fields and spinors, as well as higher dimensions. The main subtlety here is that for massive fields null infinity is not the correct boundary data to consider, and instead one would choose another ``slicing'', such as hyperboloid slicing that can resolves the field behaviour at timelike infinity $i^+$ \cite{prabhu2022infrared}. Furthermore, even in flat space, in higher even-dimensional cases the causal propagator contains higher distributional derivatives, while in odd-dimensional cases strong Huygens' principle is violated (see, e.g., \cite{Tjoa2021harvestingcomm}) despite being conformally coupled. Different spins also have different scaling behaviour for Hadamard states \cite{Takagi1986noise}. It would be interesting to see how the boundary reconstruction works out explicitly.}

Last but not least, although we have made use only of the properties of ordinary free QFT in curved spacetime, these ideas should in principle carry over to the asymptotic quantization of gravity~\cite{Ashtekar1981asymptotic,Ashtekar1981radiative,Ashtekar2018infraredissues}, and provide a new direction to explore the key differences arising from the nature of the gravitational field ({see for instance~\cite{Raju:2021lwh,Laddha:2020kvp,Chowdhury:2022wcv}}).{ We leave these lines of investigations for the future.}

\section*{Acknowledgment}

The authors thank Gerardo Garc\'ia-Moreno for pointing out some aspects of the Cauchy problem related to this setup. E.T. acknowledges generous support of Mike and Ophelia Lazaridis Fellowship.  F.G. is funded from the Natural Sciences and Engineering Research Council of Canada (NSERC) via a Vanier Canada Graduate Scholarship. This work was also partially supported by NSERC and partially by the Perimeter Institute for Theoretical Physics. Research at Perimeter Institute is supported in part by the Government of Canada through the Department of Innovation, Science and Economic Development Canada and by the Province of Ontario through the Ministry of Colleges and Universities. Perimeter Institute, Institute for Quantum Computing and the University of Waterloo are situated on the Haldimand Tract, land that was promised to the Haudenosaunee of the Six Nations of the Grand River, and is within the territory of the Neutral, Anishnawbe, and Haudenosaunee peoples.

\appendix

\section{Symplectic smearing}
\label{appendix: symplectic-smearing-Wald}

Here we reproduce, for completeness, a few results (from e.g.~\cite{wald1994quantum} Lemma 3.2.1) on the symplectic smearing \eqref{eq: symplectic smearing} and the causal propagator. First, we have the claim that \eqref{eq: symplectic smearing} is equivalent to \eqref{eq: ordinary smearing}, i.e.
\begin{equation}
    \hat{\phi}(f) \coloneqq \sigma(Ef,\hat{\phi})=\int \dd V\, f(\sx)\hat{\phi}(\sx) \,.
\end{equation}
To see this, {we can consider more generally the differential operator $P=\nabla_a\nabla^a + V\openone$, where $V\in C^\infty(\M)$ and the Klein-Gordon operator is when $V = -m^2 -\xi R$.} note that since $f(\sx)$ is compactly supported and since $\M$ is globally hyperbolic $\M \cong \R\times\Sigma_t$, there are $t_1,t_2\in\R$ such that $f=0$ for $t \notin [t_1,t_2]$. Moreover, by definition of advanced propagator $P\circ E^-f=f$, so for any $\phi\in \Sol_\R(\M)$ so that $P\phi=0$, we have
\begin{align}
    \phi(f) &= \int \dd V\,
    \phi(\sx)f(\sx) \notag\\
    &= \int_{t\in[t_1,t_2]}
    \hspace{-0.8cm}\dd V\, \phi(\sx) (P\circ E^-f)(\sx)\,\notag\\
    &=\int_{t\in[t_1,t_2]} \hspace{-0.8cm}
    \dd V\, \left[\phi\nabla^a\nabla_a (E^-f) +\phi V E^-f\right]\,.
    \label{eq: Sym smear derivation}
\end{align}

Now we need to do integration by parts. We will do this really carefully since the minus sign is a cause of confusion. We first write $\dd V = \sqrt{-g}\,\dd t\,\dd^3\bx$ where $t$ is the coordinate time associated to the foliation of $\M$, and let $\dd \Sigma = \sqrt{h}\,\dd^3\bx$ be induced 3-volume element on the spacelike surfaces $\Sigma_t$. Then we have
\begin{align}
    \phi(f) 
    &= \int_{t=t_2} \hspace{-0.35cm}
    \dd\Sigma\,
    {(-t^a)}\left[ \phi\nabla_a (E^-f)- (E^-f) \nabla_a\phi \right]\notag\\
    &-\int_{t=t_1} \hspace{-0.35cm}
    \dd\Sigma\,{(-t^a)}\left[ \phi\nabla_a (E^-f)- (E^-f) \nabla_a\phi \right]\notag\\
    &= \int_{t=t_1} \hspace{-0.35cm}
    \dd\Sigma\,{(-t^a)}\left[ (E^-f) \nabla_a\phi -\phi\nabla_a (E^-f) \right]\,,
    \label{eq: Sym smear derivation-2}
\end{align}
where $t^a$ is the future-directed unit normal vector (i.e., $t^a = \partial_t\to (1,0,0,0)$ in the adapted coordinates). The second equality follows from the fact that the smeared advanced propagator $E^-f$ {and its derivatives} vanish on $\Sigma_{t_2}$ due to $\supp(E^-f)\subseteq J^-(\supp f)$.

Using similar reasoning for the smeared retarded propagator, we also have that $E^+f$ {and its derivatives} vanish at $t_1$, so we are free replace $E^-$ in the final equality of Eq.~\eqref{eq: Sym smear derivation-2} with the causal propagator $E = E^--E^+$. Finally, by writing the directed 3-volume element as $\dd\Sigma^a\coloneqq -t^a\dd \Sigma$, so that the volume element is \textit{{past-directed}} (see, e.g., \cite{Poisson:2009pwt}), and using the definition of symplectic form \eqref{eq: symplectic form}, we get
\begin{align}
    \phi(f) 
    &= \int_{t=t_1} \hspace{-0.35cm}
    \dd\Sigma^a \,\left[(Ef) \nabla_a\phi-\phi\nabla_a (Ef) \right]\notag\\
    &= \sigma(Ef,\phi)\,,
    \label{eq: Sym smear derivation-3}
\end{align}
as desired. Hence the symplectically smeared field operator reads $\hat\phi(f) = \sigma(Ef,\hat\phi)$.  Note that as an immediate consequence of this calculation we have
\begin{align}
    \sigma(Ef,Eg) = E(f,g)
\end{align}
simply by setting $\phi(\sx)=(Eg)(\sx)$ into Eq.~\eqref{eq: Sym smear derivation-3}. 

We close this section by commenting on some issues regarding convention which can cause some confusion. In general relativity, often the convention for directed volume element is one in which it is \textit{future-directed}: that is, {$\dd\tilde{\Sigma}^a= t^a\dd\Sigma = -\dd\Sigma^a$}. In this convention, one would keep the ordering in Eq.~\eqref{eq: Sym smear derivation-2} and write the symplectic smearing as
\begin{align}
    \sigma(Ef,\phi) =  \int_{\Sigma_{t_1}}\!\!\!\dd\tilde\Sigma^a\left[ \phi\nabla_a(Ef)-(Ef)\nabla_a\phi\right]\,.
\end{align}
All we have done here is to absorb the minus sign into the integration measure. This ``freedom'' is somewhat confusing because in some cases, some authors may want to write Eq.~\eqref{eq: symplectic-smearing-alt}  ``without tilde'': in this case, the new symplectic form reads
\begin{align}
    {\sigma}'(\phi_1,\phi_2) &= \int_{\Sigma_{t_1}}\!\!\!\dd \Sigma^a\left[ \phi_2\nabla_a\phi_1-\phi_1\nabla_a\phi_2\right]\,,
    \label{eq: symplectic-smearing-alt}
\end{align}
which implies that $\sigma'=-\sigma$. In this case, by antisymmetry we have  $\sigma'(Eg,Ef) = -\sigma(Eg,Ef) = E(f,g)$. The symplectic smearing is also now defined to be $\phi(f) = -\sigma'(Ef,\phi) = \sigma'(\phi,Ef)$. Crucially, those who adopt $\sigma'$ as the symplectic form \textit{and} claim that $\sigma'(E'f,E'g) = E'(f,g)$, they will have $E'=-E$, the \textit{retarded-minus-advanced propagator}. 

Whichever convention is used, one should be consistent and one easy way to check this is as follows:
\begin{enumerate}[label=(\arabic*),leftmargin=*]
    \item Set the spacetime to be Minkowski space and fix whatever convention for $E$ and $\sigma$;
    \item Pick two functions $f,g$ and compute $Ef,Eg$, $E(f,g)$, and $\sigma(Ef,Eg)$ in the chosen convention;
    \item Using canonical quantization \cite{Tjoa2021harvestingcomm,Causality2015Eduardo}, compute $\braket{[\hat\phi(\sx),\hat\phi(\sy)]}=-\ii(\mathsf{W}(\sx,\sy)-\mathsf{W}(\sy,\sx))$, where $\mathsf{W}(\sx,\sy)$ is the unsmeared Wightman function. The standard definition is that  $\braket{[\hat\phi(\sx),\hat\phi(\sy)]}=\ii G(\sx,\sx')$, where $G(\sx,\sy)$ is the Pauli-Jordan distribution \cite{birrell1984quantum}; 
    \item Match the conventions and find the relationship between $\sigma(Ef,Eg),E(f,g)$ and $G(f,g)$ (smeared Pauli-Jordan distribution). 
\end{enumerate}
In Minkowski space we can be very explicit by choosing specific $f,g$ (even ``strongly supported'' functions like Gaussians will work). Our convention gives $\sigma(Ef,Eg) = E(f,g) = G(f,g)$ with $\hat\phi(f) = \sigma(Ef,
\hat\phi)$.

\section{BMS symmetries at $\skri^+$}
\label{appendix: BMS}

Below we briefly review some basic concepts of BMS symmetries at $\skri^+$ and its relationship as asymptotic symmetries of the bulk spacetime $\M$. It will be convenient (since we have run out of letters/symbols) to use the notation $C^\infty(\mathcal{N})$ to be the space of smooth functions on some manifold $\mathcal{N}$,  $\mathfrak{X}(\mathcal{N})$ to be the set of vector fields on $\mathcal{N}$.

\subsection{BMS group}

Recalling the definitions in Section~\ref{sec: AQFT-null}, we see that there is an inherent freedom in the definition of null infinity for an asymptotically flat spacetime: namely, the freedom to rescale the conformal factor $\Omega>0$ in a neighbourhood of $\skri^+$ by another smooth positive factor $\lambda>0$: $\Omega\to\lambda\Omega$. Under such a transformation the triple $(\skri^+, h\coloneqq g\bigr|_{\skri^+},n^a\coloneqq \tilde{\nabla}^a\Omega)$ transforms as
\begin{equation}
    \left(\skri^+, h,n\right)\longrightarrow\left(\skri^+,\lambda^2h,\lambda^{-1}n\right)\,.
\end{equation}
Thus null infinity is really the set of equivalence classes, $C=[\left(\skri^+, h,n\right)]$, of all such triples and there is in general no preferred choice or representative within a class. Moreover, null infinity is \textit{universal} in the sense that given any two equivalence classes $C_1,C_2$ with representatives $(\skri^+_1, h_1,n_1)$ and $(\skri^+_2, h_2,n_2)$, there is a diffeomorphism $\gamma:\skri^+_1\to\skri^+_2$ such that
\begin{equation}
    \left(\skri^+_2, h_2,n_2\right)=\left(\gamma(\skri^+_1),\gamma^*h_1,\gamma^*n_1\right)\,.
    \label{eq: skri-preserving}
\end{equation}
It is this freedom that allows one to transform to a Bondi frame \eqref{eq: metric-null}
\begin{align}
    \hspace{0.2cm} h_\mathcal{B} \coloneqq +2\dd u\,\dd\Omega + \gamma_{S^2}\,,
\end{align}
where $\gamma_{S^2}$ is the usual metric on the 2-sphere (not to be confused with the diffeomorphism $\gamma$) and as well $u$ is the affine parameter of the null generators $n^a \coloneqq \partial_u$.  

The diffeomorphisms $\gamma$ which preserve the equivalence classes of $\skri^+$ in the sense of \eqref{eq: skri-preserving} comprise the {Bondi--Metzner--Sachs}~\cite{bondi1962gravitational,sachs1962gravitational} group $\mathsf{BMS}_4(\skri^+)$. In other words, for any $\gamma\in\mathsf{BMS}_4(\skri^+)\subset\mathsf{Diff}(\skri^+)$ and any equivalence class $C$ with representative $\left(\skri^+, h,n\right)$ we have
\begin{equation}
    \label{eq: BMS as conf_trans}
    \left(\gamma(\skri^+),\gamma^*h,\gamma^*n \right)=\left(\skri^+,\lambda^2h,\lambda^{-1}n\right)\,.
\end{equation}
Clearly \eqref{eq: BMS as conf_trans} is independent of the representations chosen. Importantly this statement is equivalent to the following~\cite{wald2010general}: Given a one-parameter family of diffeomorphisms $\gamma_t$ generated by a vector $\tilde{\xi}$ on $\skri^+$, $\tilde{\xi}$ can be smoothly extended (not uniquely) to a vector field $\xi$ in $\M$ (for some neighbourhood of $\skri^+$) such that $\Omega^2\mathcal{L}_\xi g\to0$ in the limit to $\skri^+$.

In order to see that this definition leads to a conformal rescaling of the metric at $\skri^+$, we note that
\begin{equation}
    \Omega^2\mathcal{L}_\xi g_{ab}= \mathcal{L}_\xi \hat{g}_{ab}-2\Omega^{-1} n_c \xi^c \hat{g}_{ab}\,.
\end{equation}
Since the left hand side and the first term on the right hand side are smooth in the limit to $\skri^+$ this implies $\alpha(\xi)\coloneqq \Omega^{-1}n_c \xi^c$ is also smooth. Therefore $\Omega^2\mathcal{L}_\xi g|_{\skri^+}=0$ implies that the conformal Killing equation
\begin{align}
     \mathcal{L}_\xi \hat{g}_{ab}=2\alpha(\xi) \hat{g}_{ab}\,.
\end{align}
This preserves the null condition $n^an^b \mathcal{L}_\xi \hat{g}_{ab}={\cal O}(\Omega^{2})$. 

{Moreover, if we fix a Bondi frame $\tilde{\nabla}_{a}n_b=0$, the twist of $n_a$ vanishes, $\nabla_{[a} n_{b]}=0$, so} we also have $\mathcal{L}_\xi n^a=-\alpha(\xi) n^a$ {and ${\cal L}_n \alpha(\xi)=0$~\cite{Flanagan:2019vbl}}. By pulling back to $\skri^+$, we obtain the \textit{asymptotic symmetries} of the bulk manifold $\M$: \begin{subequations}
\begin{align}
    \mathcal{L}_{\tilde{\xi}} \gamma_{S^2} = 2\alpha(\tilde{\xi}) \gamma_{S^2}\,, \label{eq: asymp symm conf g} 
    \\
    \mathcal{L}_{\tilde{\xi}} n^a = -\alpha(\tilde{\xi}) n^a\,. 
    \label{eq: asymp symm conf n} 
\end{align}
\end{subequations}
Note that at $\skri^+$ we have $\tilde\xi=\xi$ so we will drop the tilde whenever it is clear from the context. Thus we see that these reproduce the infinitesimal action of $\mathsf{BMS}_4(\skri^+)$ (see e.g. \cite{Flanagan:2019vbl}). 

The general solution to \eqref{eq: asymp symm conf g} and \eqref{eq: asymp symm conf n} {with  ${\cal L}_n \alpha(\xi)=0$} is given by the vector field $\xi\in \VS(\skri^+)$ of the form
\begin{align}
    \xi(f,Y) = \left(f+\frac{1}{2} u\, D_A Y^A\right)n +Y\,,
\end{align}
where $n\in \VS(\skri^+)$, $Y\in \VS(S^2)$ and $f\in C^\infty(S^2)$ and
\begin{align}
    {\cal L}_n f=0={\cal L}_n Y\,\quad \mathcal{L}_Y \gamma_{S^2}=D_AY^A\gamma_{S^2}\,.
\end{align}
Note that the metric on 2-sphere $\gamma_{S^2} = \gamma_{AB}\dd x^A\dd x^B$, where $x^A$ are coordinates for $S^2$, and $\gamma_{AB}$ can be used to raise indices $A,B,C\dots$ with its associated covariant derivative $D_A$.

The vector fields $\xi(f,0) = f n$ are known as \emph{supertranslations}: they are parametrized by smooth functions $f$ on the 2-spheres and they form a ideal
of the BMS algebra $\mathfrak{bms}_4$. The smooth conformal Killing vectors of the two-sphere, $Y\in \VS(S^2)$, generate the Lorentz algebra---but there is generically no preferred Lorentz subgroup. Therefore the structure of the BMS group generated by these asymptotic Killing vectors is a semi-direct product $\mathsf{BMS}_4=SO^+(3,1)\ltimes C^\infty({S}^2)$.

We review these asymptotic symmetries in a more direct manner below.

\subsection{Asymptotic symmetries of metric}
The metric of any asymptotically flat spacetime can be written in Bondi-Sachs coordinates~\cite{bondi1962gravitational,sachs1962gravitational} 
\begin{align}\label{eq: Bondi metric}
    \hspace{-0.2cm}
    \dd s^2&= -U\dd u^2-2e^{2\beta}\dd u\dd r\notag\\
    &+g_{AB}\left(\dd x^A+\frac{1}{2}U^A\dd u\right)\left(\dd x^B+\frac{1}{2}U^B\dd u\right)\,,
\end{align}
where $\det(g_{AB})=r^4\det(\gamma_{AB})$. 

Now as a consequence of the assumptions in Sec.~\ref{sec: AQFT-null}, the large-$r$ expansion takes the form (see e.g. \cite{strominger2018lectures})
\begin{subequations}
\begin{align}
    U       &=1-\frac{2m_{\cal B}}{r}+{\cal O}(r^{-2})\,,\\
    \beta   &={\cal O}(r^{-2})\,,\\
    U_A     &=\frac{1}{r^2}D^BC_{AB}+{\cal O}(r^{-3})\,,\\
    g_{AB}  &=r^2\gamma_{AB}+ r C_{AB}+{\cal O}({r^0})\,.
\end{align}
\label{eq: asympt fall-off}
\end{subequations}
Here, $m_{\cal B}$ is the Bondi mass aspect, $C_{AB}$ is the shear tensor\footnote{Fixing Bondi gauge/coordinates and the determinant condition $\partial_r\det(g_{AB}/r^2)=0$ implies that the shear tensor is trace-free: $ \gamma^{AB} C_{AB}=0$.}, $N_A$ is the angular momentum aspect. Together with the Bondi news tensor $N_{AB}=\partial_u C_{AB}$ (and the constraint equation for the Bondi mass coming from the Einstein equations), these form the radiative data for general relativity~\cite{Ashtekar1981radiative,strominger2018lectures}. 

Observe that by introducing $\Omega=r^{-1}$ (so $\dd r=-\Omega^{-2}\dd\Omega$) and rescaling $\tilde{g}_{\mu\nu}=\Omega^2 g_{\mu\nu}$, the fall-off conditions \eqref{eq: asympt fall-off} imply that the metric in the unphysical spacetime takes the Bondi form at $\skri^+$ where $\Omega=0$, given by Eq.~\eqref{eq: metric-null}.  

Importantly one can show, by direct computation, that the fall-off conditions are preserved by the asymptotic Killing vectors~\cite{Barnich:2009se,Barnich:2010eb}
\begin{align}
    \xi&=\left(f+\frac{1}{2}uD_AY^A\right)\partial_u +\left(Y^A-\frac{1}{r}D^Af+{\cal O}(r^{-2})\right) \partial_A\notag\\
    &+ \frac{1}{2}\big(-rD_AY^A + D^AD_A f+{\cal O}(r^{-1})\big)\partial_r \,,
\end{align}
where as before these vectors, $\xi(f,Y)$, are parameterized by the scalar functions $f\equiv f(x^A)$ and the conformal Killing vectors of the 2-sphere, $Y=Y^A\partial_A$,

whose general form is~\cite{Flanagan:2015pxa}
\begin{equation}
    Y^A\equiv Y^A(x^A) = D^A\chi_e+\epsilon^{AB}D_B\chi_m
\end{equation}
where $(D^AD_A+2)\chi_{e/m}=0$, i.e. $\chi_{e/m}$ are $\ell=1$ spherical harmonics.

In particular, $\xi(0,D^A\chi_e)$ generate boosts and $\xi(0,\epsilon^{AB}D_B\chi_m)$ generate rotations~\cite{Flanagan:2015pxa}\footnote{A generalisation $Y^A$ to non smooth solutions leads to the notation of superrotations~\cite{Barnich:2009se,Barnich:2010eb} which will not concern us here.}. While expanding $f$ in a basis of spherical harmonics one finds that the first four spherical harmonics $Y_{lm}(x^A)$ correspond to ordinary translations in the bulk ($l=0,m=0$ for time translations, $l=1,m=0,\pm 1$ for spatial translations).

\subsection{Group action at $\skri^+$}

To see an explicit representation of the group at $\skri^{+}$ we will work in a Bondi frame henceforth, and  also, fix the 2-sphere to have complex stereographic coordinates $x^A=\{z,\bar{z} \}$, where $z=\cot(\theta/2)e^{i\varphi}$. In this system the Bondi frame takes the form
\begin{align}
    h_B\coloneqq +2\dd u\, \dd\Omega + \frac{4\,\dd z\,\dd\bar{z}}{(1+z\bar{z})^2}\;.
\end{align}

Keeping the notation in~\cite{Dappiaggi2005rigorous-holo} one can show~\cite{sachs1962gravitational} that the action of the $\mathsf{BMS}_4$ group takes the following form,
\begin{align}
    u'&=K_\Lambda(z,\bar{z})\left(u+f(z,\bar{z})\right)\,,\notag\\
    z'&\coloneqq \Lambda z
    =\frac{a_\Lambda z+ b_\Lambda}{ c_\Lambda z+d_\Lambda }\,,\quad
    \bar{z}'\coloneqq \Lambda\bar{z}
    =\frac{\bar{a}_{\Lambda} \bar{z}+\bar{b}_{\Lambda} }{\bar{c}_{\Lambda} \bar{z}+\bar{d}_{\Lambda} }\,.
\end{align}
Here $\Lambda\in SO^+(3,1)$ denotes a particular proper orthochronous Lorentz transformation and
\begin{align}
    K_\Lambda(z,\bar{z})=\frac{1+|z|^2}{|a_\Lambda z+ b_\Lambda|^2+|c_\Lambda z+d_\Lambda |^2}
\end{align}
and the coefficients $(a_\Lambda,b_\Lambda,c_\Lambda,d_\Lambda)$ arise from the covering map $p:SL(2,\mathbb{C})\to SO^+(3,1)$ since $SL(2,\mathbb{C})$ is a double cover of the proper orthochronous Lorentz group $SO^+(3,1)$, i.e.,  $SL(2,\mathbb{C})/\mathbb{Z}_2\cong SO^+(3,1)$.

Notice that the choice of sign does not change any of the transformations and hence we  have the semidirect product $\mathsf{BMS}_4=SO^+(3,1)\ltimes C^\infty({S}^2)$. We see the semi-direct product structure by considering the composition of two of these transformations $(\Lambda,f),(\Lambda',f')\in SO^+(3,1)\times C^\infty({S}^2)$. This yields
\begin{subequations}
\begin{align}
    &K_{\Lambda'}(\Lambda(z,\bar{z}))K_\Lambda(z,\bar{z}) =K_{\Lambda'\cdot\Lambda}(z,\bar{z})\,,\\
    &(\Lambda',f')\circ(\Lambda,f) =(\Lambda'\cdot\Lambda,f+(K_{\Lambda^{-1}}\circ\Lambda)\cdot(f'\circ\Lambda))\,,
\end{align}
\end{subequations}
where in the second line we note that $K_{\Lambda^{-1}}\circ\Lambda =1/K_{\Lambda}$.

\subsection{BMS-invariant asymptotic scalar field theory}

In order to define the action of a one-parameter element $\gamma'_t$ of the BMS group on the space of solutions $\Sol_\R(\skri^+)$ at $\skri^+$, one considers the action of its smooth extension $\gamma_t$ into $\M$ on $\phi$ and then uses the map $\Gamma$ to project it to $\skri^+$. That is, working in a Bondi frame, for $\phi \in\Sol_\R(\M)$ and $\psi \in\Sol_\R(\skri^+)$
\begin{align}
    A_{\gamma_t'}\psi
    &\coloneqq\lim_{\skri^+}\left[(\Omega_\mathcal{B})^{-1}\gamma_t^*\phi\right]\notag\\
    &=\lim _{\skri^+}\left[ \frac{\Omega_\mathcal{B}(\gamma_t(\sx))}{\Omega_\mathcal{B}(\sx)}\right]\times \lim_{\skri^+}\left[\Omega_\mathcal{B}(\gamma_t(\sx)))^{-1}\phi(\gamma_t(\sx))\right] \notag\\
    &=K_\Lambda(z,\bar{z})^{-1}\,\psi(u,z,\bar{z})\,.
\end{align}
Here, one can show making use of the asymptotic Killing equation for $\xi_t$ the generator of $\gamma_t$, that the third line follows. Alternatively this may be seen by noting that the fields $\psi\in\Sol_\R(\skri^+)$ transform with conformal weight $-1$ under the induced conformal transformation at $\skri^+$ by $\gamma_t'$ (\textit{c.f.} \eqref{eq: BMS as conf_trans}).

The induced symplectic form $\sigma_\skri$ at $\skri^+$ given by
\begin{align}
   \sigma_\skri(\psi_1,\psi_2)=\int_{\skri^+}\!\!\! \dd u\,{\dd \gamma_{S^2}}\, (\psi_1\partial_u \psi_2-\psi_2\partial_u \psi_1)\,, 
\end{align}
is also BMS-invariant because the integration measure and the derivative respectively transform as 
\begin{align}
    \dd u\,{\dd \gamma_{S^2}}\to K_\Lambda^3\,\dd u\,{\dd \gamma_{S^2}}\,,\quad \partial_u=n^a\to K_\Lambda^{-1}n^a\,.
\end{align}
Therefore all the resulting AQFT constructions (including the induced state) are BMS-invariant. 

\bibliography{ref-holography}

\end{document}